\documentclass[12pt,amsmath,amssymb,showkeys,prb]{revtex4}
\usepackage{xcolor}
\usepackage{setspace}
\usepackage{graphicx}

\DeclareRobustCommand{\citen}[1]{%
  \begingroup
    \romannumeral-`\x 
    \setcitestyle{numbers}%
    \cite{#1}%
  \endgroup
}
\newcommand{\e}{\mathrm{e}}     
\newcommand{\dd}{\mathrm{d}}  
\newcommand{\DD}{\mathcal{D}} 
\newcommand{\ii}{\mathrm i}    

\newcommand{\rr}{{\bf r }}   
\newcommand{\kk}{{\bf k }}  
\newcommand{\RR}{{\bf R }}   

\newcommand{\lB}{l_{\rm B}} 

\newcommand{\nw}{ {n_{\rm w}} } 

\begin{document}

$\null$
\hfill {\bf July 12, 2026}
\vskip 0.3in

\begin{center}

{\Large\bf Homeostatic Noise Buffering in Biomolecular}\\

\vskip 0.3cm

{\Large\bf Condensates Hinges on Phase Multiplicity}\\

\vskip 0.3cm

{\Large\bf Modulated by Interfacial and Droplet Size Effects}\\

\vskip .5in
{\bf Jonas W{\footnotesize{\bf{ESS\'EN}}}},$^{1,\dagger}$
{\bf Tanmoy P{\footnotesize{\bf{AL}}}},$^{1,\dagger}$
{\bf Suman D{\footnotesize{\bf{AS}}}},$^{2,\dagger}$
and
{\bf Hue Sun C{\footnotesize{\bf{HAN}}}}$^{1,*}$

$\null$

$^1$Department of Biochemistry,
University of Toronto, Toronto, Ontario M5S 1A8, Canada\\
$^2$Department of Chemistry, Gandhi Institute of Technology and
Management, Visakhapatnam, Andhra Pradesh 530045, India\\

\vskip 1.3cm

%

\end{center}
$\null$\\

\noindent
$^\dagger$Contributed equally.

$\null$\\


\vskip 1.3cm

\noindent
$^*$Correspondence information:\\
{\phantom{$^\dagger$}}
Hue Sun C{\footnotesize{HAN}}.$\quad$
E-mail: {\tt huesun.chan@utoronto.ca}\\
{\phantom{$^\dagger$}}
Tel: (416)978-2697; Fax: (416)978-8548\\
{\phantom{$^\dagger$}}
Department of Biochemistry, University of Toronto,
Medical Sciences Building -- 5th Fl.,\\
{\phantom{$^\dagger$}}
1 King's College Circle, Toronto, Ontario M5S 1A8, Canada.\\

\vfill\eject

\noindent
{\large\bf Abstract}\\

\noindent
Specific mixing or demixing of molecular species is a characteristic feature
of condensed intracellular membraneless compartments. How sequence patterns 
of intrinsically disordered proteins (IDPs) fundamentally impact 
subcompartmentalization of biomolecular condensates and their role in 
buffering against concentration fluctuations are hereby addressed by modeling
liquid-liquid phase separation (LLPS) of polyampholytic sequence pairs 
using random phase approximation (RPA) polymer theory and molecular 
dynamics (MD). RPA theory predicts both binary and ternary LLPS in a 
temperature-sensitive manner.  We observe demixing underpinned by ternary 
LLPS for pairs with dissimilar sequence charge 
patterns but not for pairs with similar sequence charge patterns. 
Notably, the predicted behaviors are corroborated by MD when RPA 
is augmented with interfacial tension and/or a finite-size formalism 
commensurating with the typical small sizes of MD model systems, supporting
our stipulation that RPA theory is a useful sequence-specific modeling 
tool for 
biomolecular condensates with larger, more realistic sizes when finite-size
effects are much less significant. In principle, when the condensate size is 
sufficiently large, ternary LLPS is superior to binary LLPS in noise buffering 
because the IDP compositions of the three coexisting phases in ternary LLPS 
remain unchanged over an extended two-dimensional concentration regime, 
whereas the two coexisting phases in binary LLPS are fixed only along a 
tieline. However, when condensate sizes are sufficiently small, the buffering 
capacities of ternary versus binary LLPSs are more complex as they are 
modulated differently by finite-size effects. Biophysical ramifications of 
this interplay are discussed in view of the size diversity of natural 
biomolecular condensates.
\\



\noindent
{\large\bf Significance Statement}\\

\noindent
Biomolecular condensation offers a plausible physical mechanism---among 
others---for homeostasis because phase separation fundamentally buffers
graded input concentration signals and transforms them into switch-like 
responses entailed by the discrete separated phases. In view of the notable 
size diversity of natural multi-component biomolecular condensates 
from ``nanocondensates'' to micron-sized ``membraneless organelles'', we 
build upon theories of noise buffering formulated for idealized 
infinite-size condensates and binary phase separation to delineate 
finite-size effects on sequence-specific ternary phase separation. This is 
a critical step toward elucidating noise buffering in biologically 
realistic condensates of various sizes. Our analysis also quantifies 
shortcomings of using typical small molecular dynamics-simulated systems 
as models for larger condensates and assesses how this limitation may be 
mitigated by analytical theory.


\vfill\eject


Tremendous recent advances for more than a decade \cite{cliff2009} 
have revealed that a diversity of spatiotemporal 
compartmentalizations \cite{PappuChemRev,Kate2025} afforded by membraneless 
assemblies serve important regulatory functions in biology \cite{cliff2017}
because biochemical activities can be enhanced or inhibited through 
enriching or excluding specific biomolecules and other pertinent chemical
factors in or from these assemblies. Collectively referred to as biomolecular 
condensates \cite{rosen2021}, these assemblies are often heterogeneous, 
composing of large numbers of folded domains, intrinsically disordered 
regions (IDRs) of proteins, and nucleic acids condensed into a 
biomolecule-rich state sustained by an interplay of multiple driving 
forces \cite{PappuChemRev,TanjaRohit,LinBJ2022,mohanty2023,KITP2026}. With this
complexity in mind, significant insights can be gleaned nonetheless from 
studying liquid-liquid phase separation (LLPS) of IDRs as a major physical 
underpinning of biomolecular condensation by focusing on the effects of their 
amino acid sequences as well as environmental conditions such as temperature, 
pressure, pH, and cosolvents on how the IDRs' multivalent interactions favor
or disfavor LLPS \cite{julicher2014,biochemrev,Roland2019,HXZhouRev2024,Ginelletal2025}.

In essence, LLPS is a process that transforms, as input, a set of initial 
concentrations of 
molecular species into multiple thermodynamically coexisting phases with 
different sets of concentrations as output without incurring an energy cost. 
As such, this defining feature of LLPS compartmentalization \cite{pelkmans2016} 
is a potential mechanism for passive buffering of stochastic 
noises \cite{rosen2021,biochemrev,rosen2017,Holehouse2018} in 
cellular dynamics \cite{Zechner2020,Julicher2025,dynamics}. 
The Gibbs phase rule stipulates that the 
maximum number of equilibrium phases at a given set of environmental conditions
(constant temperature and pressure) is equal to the total number of molecular 
species. As a first step toward elucidating the physics of concentration 
buffering \cite{JacobsFrenkel,Jacobs2021,Chen2023,Jacobs2023}
in multiphase condensates \cite{KnowlesPNAS2024},
three-component models with two solute species 
plus solvent were investigated \cite{biochemrev,njp2017,Safran2021}. 
Such systems can undergo binary or ternary LLPS, i.e., with two or three 
coexisting phases \cite{njp2017}. For binary LLPS, two coexisting phases remain
constant only when the initial set of concentrations lies on a one-dimensional 
(1D) tieline in the density-density plane connecting the two phases.
Effectiveness of buffering thus hinges on whether or not the noisy 
variations align approximately with the tielines across a binary LLPS 
region \cite{Safran2021}. In contrast, for ternary LLPS, all initial 
sets of concentrations falling within a two-dimensional (2D) area 
undergo LLPS to three constant coexisting phases \cite{njp2017}. This basic 
difference suggests that ternary LLPS can lead to more effective buffering 
than binary LLPS, yet this possibility has not been addressed substantively.

Another question that warrants attention is how the size (volume, $V$) of the 
condensate affects noise buffering. Analytical theories often treat LLPS in 
the thermodynamic limit with an effective infinite $V$ in which the spatial 
extents of the interfacial regions between phases 
are negligible by comparison. While 
finite-size \cite{irback2020,irback2021,Rohit2026} and experimentally measured
interfacial \cite{HXZhouRev2024,Latham2022,MittalNatComm2024,Nott2023} effects 
were considered, theoretical investigations of noise buffering thus far have 
relied mostly on infinite-$V$ formulations with infinitely 
sharp phase boundaries \cite{Julicher2025,Safran2021}. However, experiments on 
macromolecular assemblies reveal that they span a significant range of sizes 
from small clusters in subsaturated 
solutions \cite{RohitPNAS2022,Arosio2024,Lemke2025} to micron-sized 
condensates \cite{RohitPNAS2022,Arosio2024,Lemke2025,JulieRNA2022,alberti2024}. 
With this in mind, it is critical to develop formalisms for size-dependent 
noise buffering because some rather small, diffraction-sized 
``nanocondensates'' \cite{Nano2023} of linear 
dimension $\lesssim 500$ nm (below the limit of optical resolution) capable 
of regulatory 
functions \cite{40S,NanoTau2023,SUMOMolCell2023,CisseCell2024,localenviron},
including in bacteria \cite{SWeber2020}, 
can be fluid states with distinct inside environments \cite{JulieRNA2022}. 
A better understanding of size-dependent noise buffering also bears on the 
physical interpretation of computational models of biomolecular condensates 
in the face of obvious size mismatches between the small condensates 
simulated by molecular dynamics (MD) with typical linear 
dimension $\sim 15$ -- $50$ nm
\cite{dignon18,SumanPNAS,Mpipi,Kresten,Chewetal2023,Good2024,RanaCliff2024} 
and the biological condensates that are almost always significantly larger.

Here we address these fundamental questions by analyzing noise buffering in 
ternary versus binary LLPS through comparing infinite-$V$
analytical theory against finite-size coarse-grained MD. We focus on 
equilibrium aspects while recognizing that our findings will bear on future 
studies of similar systems under biologically more realistic non-equilibrium
conditions \cite{Zechner2020,Julicher2025}.
Building upon our original studies \cite{njp2017,Pal2021}, we
take advantage of the large repertoire of computational data on overall neutral 
polyampholytes as model intrinsically disordered 
proteins (IDPs) \cite{rohit2013,kings2015,Paletal_JPCL2024}. This choice of 
model is motivated by the importance of 
electrostatics \cite{rohit2013,linPRL,singperry2017,joanJPCL2019,koby2020,BenSchuler2023,PoseyJACS2024,YHL2025,BenS2025}
as a major driving force for biomolecular LLPS \cite{biochemrev,dignon2020}. 
It is instructive as well to gain a deeper insight into how comixing or 
demixing of IDP species in condensed phases can arise from a fuzzy molecular 
recognition mechanism \cite{njp2017} underpinned by a general interaction 
potential (Couloumb and screened Coulomb in this case) in lieu of an obvious 
distinction between homotypic and heterotypic 
interactions \cite{Safran2021,PappuNatComm2023}. Nonetheless, the 
conceptual framework and methodology developed here are expected to 
be applicable to biomolecular LLPS models with more heterogeneous interaction 
schemes \cite{dignon18,SumanPNAS,Mpipi,Kresten,Chewetal2023,Analytic,koby2023}.

Utilizing field-theoretic polymer representation \cite{Edwards1965,Wang2010}, 
we apply conventional random phase approximation (RPA) theory formulated 
for infinite $V$ \cite{Ermoshkin2003,MiMB2023,WessenJCP} as 
before \cite{njp2017} to obtain 
temperature-dependent phase diagrams for pairs of polyampholytes, 
determining tielines for binary LLPS and triangular ternary phase regions. 
The predicted phase behaviors are sensitive to the polyampholytes' sequence 
charge patterns. The general trend for binary LLPS is
consistent with previous studies \cite{njp2017,Pal2021}. However, by 
considering a broad range of temperatures, ternary regions hitherto 
unseen are now revealed. In these respects, the approach taken here 
for polyampholyte pairs is similar to that adopted in recent RPA studies of 
multiphase complex coacervation involving oppositely charged 
polyelectrolytes \cite{XChen2021,XChen2023}. To assess finite-size effects,
coarse-grained MD \cite{SumanPCCP2018} is employed to simulate LLPS 
in an extensive collection of systems with different initial concentrations 
to evaluate the degree to which the binary and ternary LLPS predicted 
by infinite-$V$ RPA theory are recapitulated. The results show
that while the sequence-dependent trends of MD and RPA
are similar, interfacial effects in finite-size MD are substantial.
Different initial MD concentrations within the 
ternary region predicted by infinite-$V$ RPA theory
do not separate into three fixed phases,
leading in some cases to  a wider 
final distribution of concentrations than that entailed by binary LLPS,
underscoring the significant impact of the finite-size effects 
on noise buffering. Notably, we find that 
these finite-size phenomena can be rationalized 
physically by augmenting RPA theory with an explicit surface tension 
term \cite{Zechner2020,Latham2022,MittalNatComm2024} and/or a finite-$V$,
Gibbs ensemble-like formulation of RPA.
In both cases, conventional RPA is recovered in the infinite-$V$ limit.
Accordingly, as described below, our analysis not only offers new 
insights into the interplay of condensate 
size and phase multiplicity in noise buffering; it also affords a basis
for extrapolating from small-size MD simulations to the larger 
biomolecular condensates they aim to model.
\\



\noindent
{\large\bf Results}\\

As outlined in Materials and Methods, the formulation below was developed from
our previous analytical RPA and coarse-grained MD studies of 
biomolecular LLPS \cite{linPRL,SumanPCCP2018,MiMB2023,WessenJCP} 
particularly those involving two solute 
species \cite{njp2017,Pal2021}. Technical details are given in SI Text.
\\

\noindent
{\bf RPA Theory Predicts Ternary LLPS for Polyampholyte Pairs} 

Fig.~1 overviews our approach. To probe sequence dependence,
we study seven fully charged, overall neutral 50-bead polyampholytic sequences 
(Fig.~1a) that span a wide range of charge patterns from strictly alternating 
(sv1) to highly blocky (sv28), taken from a set of model sequences 
that have been studied extensively \cite{rohit2013,kings2015,Paletal_JPCL2024}
(all sequences studied in this work are listed in Table~S1).
We focus on six two-solute systems 
consisting of each of the first six sequences (sv1, sv10, sv15, sv20, sv23, 
sv24) pairing with sv28 (as the solute pair), representing a range of 
differences in charge pattern between the paired sequences that may be 
quantified by parameters such as $\kappa$ \cite{rohit2013} or sequence charge 
decoration (SCD) \cite{kings2015}.

We model infinite-$V$ phase behaviors by RPA theory using a field-theoretical 
representation of polyampholytes that accounts only for chain 
connectivity, Coulomb interactions, and excluded volume (Materials and Methods) 
\cite{linPRL,WessenJCP}. In particular, phase diagrams for sequence pairs are 
constructed from the RPA free energy per unit volume 
and the product of Boltzmann's constant $k_{\rm B}$ and absolute 
temperature $T$, $f_{\rm RPA}$ 
[Eq.~\ref{eq:rpa_free_energy}], using standard methods \cite{MiMB2023}. 
Examples are provided in the three-dimensional plots in Fig.~1b--d for 
the sv10-sv28, sv15-sv28, and sv23-sv28 pairs as a function of $T$ and
the two sequences' volume fractions ($\phi$s). Depending on the sequence 
pair and $T$ shown, either exclusively binary LLPS (tielines only) or 
a combination of binary and ternary LLPS (tielines and red triangle) is 
observed at a given $T$. Notably, for a given sequence
pair, ternary LLPS can happen at some $T$s but not others
(Fig.~1b). These results go beyond 
our previous study of sequence-dependent LLPS in two-solute systems with
only binary LLPS \cite{njp2017} and a related study of 
noise buffering that considered only binary LLPS \cite{Safran2021}.
As before \cite{linPRL,WessenJCP}, the reduced temperature $T^*$ 
here is $T$ in units of 
$4\pi b\epsilon_0\epsilon_{\rm r}k_{\rm B}/e^2$; i.e., $T^*=b/l_{\rm B}$ 
where $b$ is the equilibrium bond length between successive charged beads, 
$l_{\rm B}=e^2/4\pi\epsilon_0\epsilon_{\rm r}k_{\rm B}T$ is the
Bjerrum length, $e$ is protonic charge, $\epsilon_0$ and $\epsilon_{\rm r}$ 
are, respectively, vacuum and relative permittivity. 

Further infinite-$V$ sequence-dependent behaviors are given by the 
2D phase diagrams in Fig.~1e--j for all six sequence pairs.
The sv1-sv28 diagrams computed at a low $T^*$ (Fig.~1e) illustrates that 
any initial, or overall average (input) volume fractions 
$(\bar{\phi}_{\rm sv1},\bar{\phi}_{\rm sv28})$ within the triangular 
ternary region always phase separates into the three phases with fixed
volume fractions defined by the vertices of the triangular
region \cite{njp2017}. The trend of binary LLPS in Fig.~1e--j is
consistent with our previous finding that sequence pairs with a larger 
difference in charge pattern tend to demix in the condensed phase
(tielines with negative slopes) whereas sequence pairs with similar
charge patterns tend to co-mix in both dilute and condensed phases
(tielines with positive slopes) \cite{njp2017,Pal2021}. Notably, some of
these phase digrams exhibit a mixture of ternary and binary LLPS (Fig.~1e--g)
while others exhibit only binary LLPS (Fig.~1h--j). 
More extensive LLPS data for different polyampholytic sequence pairs 
involving other sv-sequences and sequences with constant $\kappa$ or 
constant SCD \cite{Paletal_JPCL2024} are provided in Figs.~S1--4.
Althought Fig.~1e--g may suggest that 
ternary LLPS is more likely for sequence pairs that demix, the 
generality of this trend awaits further investigation. This is because 
the onset of ternary LLPS depends on $T^*$ (Fig.~1b--d). It
appears at a lower $T^*$ for some but at a higher $T^*$ 
for other polyampholyte pairs (Fig.~S1--S4). The trend also does not 
hold for polyelectrolyte pairs (see below).
\\

\noindent
{\bf MD-Simulated Condensed Droplets Exhibit Substantial Interfacial Features}

We complement RPA with coarse-grained MD (Fig.~2) using the ``1/3 LJ'' 
potential \cite{SumanPCCP2018}, which includes a small 
sequence-independent attraction as part of the Lennard-Jones 
(LJ) term that also accounts for excluded volume (Materials and Methods). Such
a term is needed computationally for sv1 to undergo LLPS at a 
sufficiently high $T^*$ that allows for equilibration of 
systems involving also more strongly interacting sequences (e.g. sv28)
\cite{SumanPCCP2018,Pal2021}. Despite this minor mismatch between MD and RPA, 
comparisons between 1/3 LJ MD and RPA models have proven useful for 
physical insights into polyampholyte LLPS \cite{SumanPCCP2018,Pal2021}.

The MD snapshots in Fig.~2 exhibit features 
akin to both binary and ternary LLPS. Fig.~2a,b 
shows the transition from a fully mixed state 
(Fig.~2a) to a binary LLPS state with a sv28-rich condensed 
droplet and a sv1-rich dilute phase when $T^*$ is decreased (Fig.~2b). 
At an even lower $T^*$, the sv1-sv28 density distribution (Fig.~2c,d) 
resemble that of ternary LLPS, namely that a condensed droplet 
consisting of a sv28-rich core enveloped by a sv1-rich region 
is surrounded by a volume dilute in both sv1 and sv28.
(The sv28-rich core in some snapshots may
appears as a tube due to periodic boundary conditions; see Fig.~S5.)
In comparison, each of the snapshots in
Fig.~2e--f for the sv10-sv28, sv15-sv28, and sv23-sv28 pairs simulated
at the same $T^*$ has a more mixed condensed droplet without a very clear
separation into subcompartments, resembling to a certain extent binary LLPS 
with a relatively mixed single condensed phase and a dilute phase, though sv28 
is still more concentrated in the core of the condensed droplets 
for sv10-sv28 (Fig.~1e) and sv15-sv28 (Fig.~1f).
Taken together, this trend is in line with our previous
MD study \cite{Pal2021} and is qualitatively consistent with the RPA 
results in Fig.~1e--j with the aforementioned proviso that because of the
overall 1/3 LJ term in the MD potential, the two polyampholyte species 
in the MD condensed phase is unsurprisingly more mixed than predicted by RPA.

To assess the degree to which MD-simulated droplets conform to 
RPA-predicted binary and ternary LLPS, we perform MD at a low $T^*$
for an extensively varied set of initial densities (or, equivalently, 
initial concentrations or volume fractions). 
The resulting MD profiles in Fig.~S6 for sv28 (blue curves) 
and its pairing sequence (curves in a 
different color) are the bead density $\rho(z)$, i.e., number of 
beads per unit volume), where $z$ is the coordinate of
the simulation box's elongated dimension (Materials and Methods). If 
$b$ is equated with the {\it trans} polypeptide C$_\alpha$--C$_\alpha$ virtual 
bond length $\approx 3.8$~\AA, the condensed droplets in 
Fig.~S6 defined by a region with $\rho$ substantially larger than zero 
span a linear
dimension of $\approx 15$ nm. In line with the trend in Fig.~2, for sv1-sv28, 
sv10-sv28, and sv15-sv28, which have dissimilar sequence charge patterns 
within each pair, the two sequence species in the condensed droplet demix 
appreciably for a range of initial overall densities, with a
discernible peak-sv28 region flanked by two peak regions for the 
pairing sequence. These features are ternary LLPS-like, showcasing two 
distinct condensed regions 
and a polyampholyte-dilute region. However, unlike LLPS 
with sharp phase boundaries stipulated by infinite-$V$ RPA theory (Fig.~1), 
substantial interfacial effects are manifested in the MD-simulated sv1-sv28, 
sv10-sv28 and sv15-sv28 droplets by the sizable population 
overlaps of the two sequence species (Fig.~S6). 
Indeed, these population overlaps increase as the sequence charge patterns 
of the pair become less dissimilar (sv1-sv28 $\rightarrow$ sv10-sv28 
$\rightarrow$ sv15-sv28).
When the sequence charge patterns become even more similar (sv20-sv28, 
sv23-28, and sv24-sv28), the condensed populations of the two sequence 
species in the pair are even more mixed, though some degree of demixing
can still be discerned for sv20-sv28 and sv23-sv28.
\\

\noindent
{\bf Typical MD-Simulated Droplets Deviate from Predicted Ternary LLPS}

To evaluate the similarity between MD-simulated droplets and RPA-predicted 
binary or ternary LLPS, we
first assess how polyampholyte density profiles behave upon variation in 
initial (overall) polyampholyte densities, focusing on the sv10-sv28, 
sv15-sv28, and sv23-sv28 pairs as examples 
(Fig.~3). As noted, most of the MD profiles for sv10-sv28 and 
sv15-sv28 in Fig.~S6 exhibit ostensibly ternary LLPS features with
three discernibly distinct density regimes. But to be in full compliance with
ternary LLPS into three fixed phases, the relative populations of the two 
species should be a constant in the core sv28-rich region
and another constant in the sv10- or sv15-rich flanking regions irrespective 
of the initial overall densities (Fig.~1f,g). 
This is not the case for the sv10-sv28 and sv15-sv28 MD droplets (Fig.~3a) 
because the relative sequence populations in the 
sv28-rich core and the sv10- or sv15-rich regimes vary substantially 
with initial polyampholyte densities. For sv23-sv28, 
Fig.~3a indicates that relative sequence populations also vary with initial 
densities but the sv23-sv28 MD droplet lacks a prominent sv23-rich regime
and is quite well mixed instead. In this respect, the sv23-sv28 MD system
in Fig.~S6 and Fig.~3 resembles binary LLPS into a dilute 
and an essentially homogeneous condensed phase (Fig.~1i).
\\

\noindent
{\bf MD Droplets are Describable by
Theory with an Explicit Interfacial Term}

To account for the salient 
features of MD-simulated droplets and finite-size condensates in general,
we develop a phenomenological model that extends 
$f_{\rm RPA}$ for spatially uniform component densities (in an 
effectively infinite $V$) to a phase-field functional \cite{PhaseField1996} 
$F[\lbrace \rho_{p}(\rr) \rbrace]$ that allows for local density 
variation with an augmented term for interfacial effects, viz.,
\begin{equation}
Vf_{\rm RPA} 
\rightarrow
F[\lbrace \rho_{p}(\rr) \rbrace] = 
\int \dd \rr \left[ \sum_{p,q=1}^M \frac{\xi_{p,q}}{2} \;
\nabla\rho_p(\rr) 
\cdot 
\nabla\rho_q(\rr) + f_{\rm RPA}(\lbrace \rho_p(\rr) \rbrace) \right]
\; ,
\label{F_eq}
\end{equation}
where the original $f_{\rm RPA} = f_{\rm RPA}(\{\bar{\rho}_p\})$ depends on 
uniform densities $\bar{\rho}_p$; $p,q$ are
component labels, and $V=\int \dd \rr$ can be finite.
The term with the $\xi_{p,q}$ interface
coefficients penalizes density gradients of individual components 
($\xi_{p,p}>0$ for $p=q$) and provides for possible crosstalks between 
density gradients of different components ($p\ne q$). For spatially uniform
densities, this term vanishes and $F$ reverts to $V f_{\rm RPA}$.

We arrive at a set of empirical 
$\xi_{p,q}$s by using the surface tension $\gamma_{\rm MD}$s 
computed by Devajaran et al. for the single-sequence condensed phase 
of 14 50-bead sv-type ``E-K'' sequences with a potential similar 
though not identical to ours \cite{MittalNatComm2024}.
Combining an approximate linear dependence of $\gamma_{\rm MD}$ 
on SCD for sv-type sequences (Fig.~S7) and
the relationship between single-sequence surface tension and $p=q$ 
interface coefficient (SI text) yields
\begin{equation}
    \xi_{p,p} = 246 \left[0.368 - 0.020 ({\rm SCD}) \right]b^8  \, 
\end{equation}
in the phase-field model for our MD droplets. We further
set the cross-sequence coefficients 
\begin{equation}
\xi_{p,q} = \sqrt{\xi_{p,p} \xi_{q,q}} 
\frac{\min(|\mathrm{SCD}_p|,|\mathrm{SCD}_q|)}{\max(|\mathrm{SCD}_p|,|\mathrm{SCD}_q|)} \, 
\end{equation}
because they reduce to $\xi_{p,p}$ when $p=q$ as they should and,
retrospectively, produce reasonable interfaces in systems of
sv sequence pairs.

Equilibrium phase-field density distributions 
are numerically solved as the steady state of a Cahn-Hilliard-type 
process \cite{Cahn1958} in a finite-size box (SI Text). 
Phase-field $\rho_p(\rr)$ is a function of $\rr=(x,y,z)$; but to
facilitate comparison with 1D MD profiles (Fig.~S6), most of 
our phase-field results are obtained by assuming $\rho(\rr)=\rho(z)$
and thus $\phi=\phi(z)$ is independent of $x,y$ (Fig.~3b,c), while noting
that a 2D implementation may be necessary
when the core phase is so small that subcompartmentalization is 
difficult to discern from a 1D profile (Fig.~S8). In 1D implementations,
system size is defined by the length $L$ along the 
$z$-axis. When $L$ is small, phase-field results in Fig.~3b show 
substantial variations in sequence populations depending on 
initial $\bar{\phi}$s. These variations, which are prominent in the 
sv28-rich cores of sv10-sv28, sv15-sv28 as well as the entire 
sv23-sv28 droplet, are quite similar to the trends seen for the 
MD systems (Fig.~3a). Notably, as system size increases,
the phase-field-modeled variations in the sv10-sv28 and sv15-sv28 systems 
decrease but the variation in the sv23-sv28 system persist. Ultimately,
in the infinite-$V$ ($L\rightarrow \infty$) limit, the
variations vanish for sv10-sv28 and sv15-sv28 (Fig.~3c), 
recovering the original RPA-predicted ternary LLPS for sv10-sv28 and 
sv15-sv28 but binary LLPS for sv23-sv28 (Fig.~1f,g,i). Taking the 
similarity between finite-size MD and phase-field behaviors (Fig.~3a,b) 
together with the infinite-$V$ phase-field limit (Fig.~3c), one can 
reasonably expect that the deviation from ternary LLPS properties in MD 
droplets with three discernibly distinct density regimes is a finite 
size effect that should diminish with increasing size of the MD system.
Although a thorough assessment of this expectation by simulating 
MD systems of much larger sizes is not currently feasible, we took an 
exploratory step by performing MD of 1,000-chain sv10-sv28 systems with initial
(overall) sv10/sv28 density ratios 3/7 ($n_{\rm sv10}=300$ sv10 chains, 
$n_{\rm sv28}=700$ sv28 chains) and 7/3 
[$(n_{\rm sv10},n_{\rm sv28})=(700,300)$]. Consistent with the expected trend,
comparison with the corresponding 500-chain systems in Fig.~S9 indicates that
when the initial sv10/sv28 density ratio changes from 3/7 to 7/3, the
sv10-sv28 population ratio $\rho_{\rm sv28}/\rho_{\rm sv10}$ in the 
sv28-rich core 
changes from $\rho_{\rm sv28}/\rho_{\rm sv10}\approx 3.4$ to $0.6$ 
for the 1,000-chain system, which is less than the 
changes from $\approx 5.4$ to $0.8$ 
for the 500-chain system,
and that $\rho_{\rm sv28}/\rho_{\rm sv10}$
is notably more uniform in the 1,000-chain than in the 500-chain 
7/3 MD system (Fig.~S9, right panel).
\\

\noindent
{\bf Noise Buffering by Finite-Size Condensed Droplets}

Moving beyond 1D density profiles, Fig.~4 shows 2D equilibrium density 
distributions of MD-simulated sequence pairs. For the
sv1-sv28, sv10-sv28, and sv15-sv28 pairs with three discernibly distinct 
density regimes and thus ternary LLPS-like (Fig.~S6), variation in 
initial densities results in highly diffused final distributions (Fig.~4, top 
row), far from the defining property of ternary LLPS that only three 
phases at the vertices of a triangular region are populated. Here, only
a slightly higher population (a small red area) near the vertical 
axes for sv1-sv28 and sv10-sv28 hints at a connection
to ternary LLPS. Indeed, these 
distributions are almost indistinguishable from that of the sv23-sv28 pair
predicted to undergo binary LLPS. Representative 
distributions for a single initial density are also provided in 
Fig.~4 (all except the top row), exhibiting highly diffused output densities
for a single input instead of separating into only two or only 
three densities as stipulated by infinite-$V$ RPA. Thus,
a dispersed, or noisy, output can be generated by a small 
condensate even in the absence of input noise.  Indeed, when compared to
RPA 2D ternary LLPS phase diagrams (Fig.~1e--g), the sv1-sv28, 
sv10-sv28, and sv15-sv28 MD distributions in Fig.~4 are much broader
although the existence of some essentially unpopulated areas below the main 
populated diagonal is akin to RPA-predicted phase behavior. The same features 
are seen for all sequence pairs and initial 
densities we studied (Figs.~S10--S12). 

A semi-quantitative account of the diffused MD-simulated 2D density 
distributions from varying initial densities (top row of Fig.~4) 
is afforded by phase-field theory (Fig.~5a). The similarity between the
MD and phase-field 2D density maps supports the contention that the 
diffused density (or volume fraction) distributions in MD---especially for 
sequence pairs with three discernibly distinct density regimes---are caused by 
finite-size interfacial effects. 

We now utilize this connection between phase-field theory and MD to 
explore how noise buffering depends on droplet size.
While dilute-phase concentration buffering and noise reduction are distinct 
properties of condensates especially when operating under non-equilibrium 
conditions \cite{Julicher2025}, our focus, similar to that of
Ref.~\citen{Safran2021}, is on equilibrium systems and how noisy input 
concentrations may be ``buffered'' or transformed into stable, well-defined 
concentrations in condensed phases.
To this end, we consider a normal distribution of initial $\bar{\phi}$s
on the $\phi_i$--$\phi_j$ plane (cluster 
of yellow stars in Fig.~5b). Because this noisy set lies entirely 
within the triangular ternary LLPS regimes of sv10-sv28 
and sv15-sv28 (Fig.~5c), perfect buffering can be achieved with 
only three populated vertex phases in the infinite-$V$ 
($L\rightarrow\infty$) limit. Phase-field modeling is performed for 
$L=128b$---which is small relative to real biomolecular
condensates but slightly larger than the $L=64b$ fitted to the MD
data in Fig.~3---to probe how volume fraction 
distribution varies with system size.  At $L=128b$, the phase-field 
2D distributions in Fig.~5b for sv10-sv28 and sv15-sv28 have nonzero 
populations between the positions of two pairs of vertices, unlike the
zero population at these positions for $L\rightarrow\infty$. 
Nonetheless, the red spots on these maps indicate that the ternary vertex 
positions are most populated. 
The phase-field distribution for sv23-sv28 in Fig.~5b is also 
diffused with nonzero population in the $L\rightarrow\infty$ binary LLPS
region; but kinship with infinite-$V$ behavior is indicated by the 
enhanced populations (red areas in the rightmost sv23-sv28 panel) near 
the origin and a range of positions coinciding with
the binodal phase boundary in Fig.~5c. The 1D profiles 
corresponding to the 2D maps in Fig.~5b are provided in 
Fig.~5d. The diffused distributions in Fig.~5b echo those displayed in 
the MD density maps for a single input density (Fig.~4, bottom three rows), 
underscoring the shared finite-size origin of these common features. 
Intuitively, that
the phase-field distributions in Fig.~5b are less diffused than the MD 
distributions in Fig.~4 is expected owing to the aforementioned larger 
effective phase-field system size used in Fig.~5.
\\

\noindent
{\bf Noise Buffering by Ternary LLPS Improves with Droplet Volume}

We next leverage $L$-dependent phase-field results 
to monitor how the efficacy of LLPS-modulated 
buffering of noise in input $\bar{\phi}$s
depends on droplet size (Fig.~6).
Two complementary measures are analyzed.
The first is based on the $P(\phi_1,\phi_2)$ 
distributions (Fig.~5b)
constructed by binning the equilibrium phase-field
$(\phi_i, \phi_{\rm sv28})$ 
on a $50\times 50$ grid with the initial
$\bar{\phi}_{i,{\rm sv28}}$ drawn, e.g., 
from the normal distributions (yellow stars) in Fig.~5b,c.
An $L$-dependent
information entropy-like $S(P(\phi_1,\phi_2)) \equiv 
-\sum_{\phi_1,\phi_2} P(\phi_1,\phi_2) \ln P(\phi_1,\phi_2)$ is then 
defined for a given initial $(\bar{\phi}_1,\bar{\phi}_2)$, which measures
the volume fraction dispersion, or noise, generated by LLPS from 
a single, sharp initial $(\bar{\phi}_1,\bar{\phi}_2)$ with no input noise.
Not surprisingly, $S(P)$ decreases monotonically with increasing $L$ (Fig.~6b).
To characterize the total volume fraction dispersion
(beyond sharp-input baselines) for a set of noisy, i.e., multiple, input,
$S$ is applied to the average $\langle P(\phi_1,\phi_2)\rangle$ over a 
noisy $\{(\bar{\phi}_1,\bar{\phi}_2)\}$ set of inputs as in Fig.~5b,c,
leading to $S(\langle P(\phi_1,\phi_2)\rangle)$ as
buffering efficacy for the given set of inputs,
where $\langle\dots\rangle$ denotes averaging over 
different $(\bar{\phi}_1,\bar{\phi}_2)$s.
A smaller $S(\langle P\rangle)$ signals fewer populated states in 
the final equilibrated system and therefore more suppression of input noise.
The variations of $S(\langle P\rangle)$ with $L$ 
are plotted in Fig.~6a.
As expected, the ternary LLPS-like sv10-sv28 and sv15-sv28 pairs
provide more effective 
buffering of input noise than the binary LLPS-like sv23-sv28 system 
in the infinite-$V$ 
($L\rightarrow\infty$) limit. However, by this 
$S(\langle P\rangle)$ measure, all three systems have 
similar noise buffering efficacies for small $L$. 
Indeed, Fig.~6a indicates
that the binary LLPS-like sv23-sv28 provides slightly better 
buffering than the two ternary LLPS-like s10-sv28 and 
sv15-sv28 for $L\lesssim 200b$. 
By comparison, when input noise is absence, i.e., for $S$ applied to 
one $(\bar{\phi}_1,\bar{\phi}_2)$ at a time, there is no marked difference
in trend between ternary and binary LLPS-like systems with regard to their 
decreasing output dispersion $S(P)$s with increasing $L$ (Fig.~6b).
Hence it is reasonable to conclude from the ternary-binary crossover in 
Fig.~6a at $L\approx 200b$ that ternary LLPS-like systems are more effective 
suppressors of input noise than binary LLPS-like systems only at sufficiently 
large system sizes, underscoring a complex interplay of binary versus ternary 
LLPS-like behaviors and system size in noise buffering. 

Our second measure of noise buffering is
based on the variation in equilibrated (output)
$\phi$s in the core of a 
droplet due to varied inputs. Exemplified by 
Fig.~3, effective noise buffering amounts to little or no variation 
in output irrespective of input. 
We compute standard deviations ($\sigma_\phi$s)
of phase-field $\phi$s of sequence 
species at the $z=0$ center of the droplet for different $L$s (Fig.~6c--e). 
According to this $\sigma_\phi$ meausre, the three sequence pairs have 
comparably weak noise buffering (high $\sigma_\phi$s) for small 
$L$, but the ternary 
LLPS-like sv10-sv28 and sv15-sv28 (Fig.~6c,d) systems are dramatically more
effective (i.e., have significantly lower $\sigma_\phi$s) than the binary 
LLPS-like sv23-sv28 system (Fig.~6e) for $L\gtrsim 200b$. 

Notably, both the $S(\langle P\rangle)$ (Fig.~6a) and $\sigma_\phi$
(Fig.~6c--e) measures point consistently to a threshold $L\approx 200b$ for 
highly effective noise buffering through ternary LLPS 
of the polyampholyte systems considered. If $b$ is identified as the 
{\it trans} C$_\alpha$-C$_\alpha$ virtual bond length $\approx 0.38$ nm,
$L\approx 200b$ corresponds to a condensate with
linear dimension $\sim 80$ nm. This stipulation may suggest
that two-sequence condensates can be highly effective in
buffering input noise at a level approaching that of an ideal 
(infinite-$V$) ternary LLPS system in real biological nanocondensates with 
linear dimension $\lesssim 500$ nm \cite{JulieRNA2022,alberti2024,Nano2023}.
However, phase-field predictions 
of $L$-dependent surface tensions are governed by the interfacial 
coefficients which are sensitive to the chain lengths and interactions 
of the polymer molecules in the LLPS system. Therefore, the extent to 
which this $L\sim 80$ nm threshold for effective noise buffering via
ternary LLPS is applicable beyond the present type of 50-bead polyampholytic
sequence pairs remains to be investigated.
\\



\noindent
{\bf Concentration Dispersion in MD is in line
with Finite-Size Thermal Fluctuation}

Beside interfacial effects, another physical basis for concentration 
dispersion in finite-size LLPS systems is thermal effects as fluctuations 
increase with decreasing system size. To assess this effect on
any free energy/$k_{\rm B}T$ per unit volume,
$f(\{\phi_p\})$, where $p$ is component label,
we consider a Boltzmann-weighted system described by a
partition function with total volume $V$ equally 
divided into $N_{\rm v}$ subsystems (voxels) with volume $V_{\rm v}$:
\begin{equation}
Z[\{\bar{\phi}_p\};V_{\rm v},N_{\rm v}] = 
\sum_{\{\phi_p^{(k)}=0\}}^{\{\phi_p^{(k)}=1\}}
\exp\left ( -V_{\rm v}\sum_{k=1}^{N_{\rm v}} f(\{\phi_p^{(k)}\}) \right ) 
\prod_p \delta \left ( \bar{\phi_p} - \sum_{k=1}^{N_{\rm v}} 
\phi_p^{(k)}/N_{\rm v} \right)
\label{Z_eq}
\end{equation}
where the exponent for the total Boltzmann factor is the 
sum over exponents for the $k$-labeled 
voxels; the product of $\delta$-functions constrains summations of 
voxel $\{\bar{\phi}_p^{(k)}\}$ 
to be the input $\{\bar{\phi}_p\}$; and $V=N_{\rm v}V_{\rm v}$. 
This setup is primed for capturing thermal effects in a finite-size system by
allowing $\{\phi_p\}$ with higher-than-lowest free energies to be sampled. 
However, since Eq.~\ref{Z_eq} contains no
spatial information and prescribes no interaction between voxels,
this formulation---unlike the phase-field model---does not address 
contributions from interfacial effects. As such, this system is similar 
to the Gibbs ensemble for phase equilibria \cite{Panagiotopoulos1988} 
except here we consider many voxels ($N_{\rm v}\gg 1$) 
rather than sample systems with only two or 
a small number of simulation boxes to approximate infinite-$V$ phase behavior.
The formulation in Eq.~\ref{Z_eq} is driven solely by 
the Boltzmann distribution; 
the free energy function under LLPS conditions is automatically 
recovered in the infinite-$V$ limit without needing the conventional 
consideration of chemical potential and osmotic pressure \cite{MiMB2023}, viz.,
\begin{equation}
\lim_{V,V_{\rm v}\to\infty}
\left (
-\ln Z[\{\bar{\phi}_p\};V_{\rm v},N_{\rm v}]/V
\right ) =
\sum_\alpha v_\alpha f(\{\phi_p^{\alpha}\}) \; ,
\label{V_eq}
\end{equation}
where $\alpha$ labels the separated phases, $\sum_\alpha v_\alpha =1$, and 
$\sum_\alpha v_\alpha \phi^{\alpha}_p = \bar{\phi}_p$. 
When the system is homogeneous (only one value of $\alpha$, no LLPS),
this limit reduces to $f(\{\bar{\phi}_p\})$ as it should.

Equilibrium distributions of $Z$ are obtained by Monte Carlo (MC) 
sampling $\phi^{(k)}_p$s while maintaining 
$\{\bar{\phi}_p\}$ with move acceptance probabilities 
consistent with the Boltzmann weighting in Eq.~\ref{Z_eq}, resulting
in a distribution of $\{\phi_p\}$ for the voxels.
(SI Text). As a test, we consider
$f(\phi)=\phi\ln\phi+(1-\phi)\ln(1-\phi)$ for a homogeneous 
system of beads with excluded volume but non-interacting otherwise.
The algorithm yields 
$\langle (\phi-\bar{\phi})^2\rangle\approx\bar{\phi}(1-\bar{\phi})/V_{\rm v}$
$\propto V_{\rm v}^{-1}$,
which is consistent with the general expression\cite{Mishin}
$\langle (\phi-\bar{\phi})^2\rangle=V_{\rm v}(\partial^2 f/\partial\phi^2)_T^{-1}$
for density fluctuation of any $V_{\rm v}$ in a homogeneous
system (SI Text). An example with two coexisting phases is 
provided in Fig.~7a--d for a Flory-Huggins (FH) model\cite{biochemrev} with
infinite-$V$ dilute- and condensed-phase 
volume fractions $\phi_{\rm dil}=0.1200$ and $\phi_{\rm con}=0.6684$ 
(dashed lines in Fig.~7a) determined conventionally \cite{njp2017,MiMB2023}. 
Fig.~7b--d show the 
MC distributions $P(\phi)$ 
for four finite-$V$ systems each consisting of 200 voxels, 
with initial volume fraction $\bar{\phi}$ slightly above $\phi_{\rm dil}$ (b),
in the middle of the LLPS region (c), or slightly above $\phi_{\rm con}$ (d).
Larger thermal fluctuations are seen for smaller volumes.
The spread around each $P(\phi)$ peak is
quite broad for smaller $V_{\rm v}$ but narrows with increasing $V_{\rm v}$.
In line with Eq.~\ref{V_eq}, 
when the input $\bar{\phi}$ is well within the LLPS region (Fig.~7c)
two separated $P(\phi)$ peaks at $\phi_{\rm dil}$ and 
$\phi_{\rm con}$ become individually sharper 
as $V_{\rm v}$ increases. In this LLPS (inhomogeneous) case, the variance 
in $\phi$
is dominated by the difference between $\phi_{\rm dil}$ and $\phi_{\rm con}$.
Even when thermal fluctuation is eliminated by $V_{\rm v}\rightarrow\infty$ or
$T\rightarrow 0$, $\langle (\phi-\bar{\phi})^2\rangle\rightarrow
(\bar{\phi}-\phi_{\rm dil})(\phi_{\rm con}-\bar{\phi})$ does not vanish.

The model of Eq.~\ref{Z_eq} is then applied to
the RPA models for the sequence pairs 
in Fig.~1f,g,i with MC model 
$V=$ $10^6 b^3$, $5\times 10^6 b^3$, $10^7 b^3$,
and $V_{\rm v}=V/200$. Similar to the $P(\phi)$s in Fig.~7b--d, the 
$P(\phi_1,\phi_2)$s in Fig.~7e--g are highly dispersed
at small $V_{\rm v}$ (top panels), resembling those
of the MD snapshots (Fig.~4). As 
$V_{\rm v}$ increases, the MC distributions sharpen and 
approach those for infinite-$V$ ternary (Fig.~7e,f) or binary (Fig.~7g) 
LLPS (bottom panels). A similar trend is seen in Fig.~S13 for a previous FH
model of ternary LLPS~\cite{njp2017}.
For the two ternary LLPS-like systems at small $V_{\rm v}$ 
(Fig.~7e,f, top), the regions between the
sv28-rich $\approx (0,0.2)$ areas and the
sv10- or sv15-rich $\approx (0.1,0)$ areas are
substantially populated. The region along the horizontal axis
between the sv10- or sv15-rich region and
the dilute population peak (red dot near the origin)
is also appreciably populated. In contrast, the region along the vertical
axis between the sv28-rich region and the origin is not populated.
It is noteworthy that these MC features are consistent with the MD density 
distributions (Fig.~4) for droplets with an sv28-rich core
and an sv10- or sv15-rich shell (Fig.~2 and Fig.~S6).

A rigorous match between the system sizes of MC 
voxel-ensemble models (Eq.~\ref{Z_eq}) and MD is not 
warranted because the MC models lack interfacial effects. Nonetheless,
a semi-quantitative connection is instructive.
To accommodate the $2.5\times 10^4$ beads of size $\approx b^3$ in an 
MD system of 500 50-bead chains (Fig.~4), the required volume $V$ ($/b^3$)
of an MC system similar to those in Fig.~7e--f with total polymer volume 
fraction $2\times 0.0347=0.0694$ is 
$V\approx 2.5\times 10^4/0.0694\approx 3.6\times 10^5$, which is $\approx 1/3$
of $V=10^6 b^3$ for the top panels of Fig.~7e--g (see SI Text for
further analysis). In light of this rough estimate, the similar dispersions 
seen in these results by MC and those by MD
in Fig.~4 indicate that a part of the MD density dispersions 
likely originates from finite-$V$ thermal fluctuations.
By the same token, since the MD droplets have $\rho \approx 0.7b^{-3}$ 
(Fig.~3a), the significantly reduced dispersions exhibited by the bottom 
panels of Fig.~7e--g for $V=10^7 b^3$ suggests that thermal effects cease to entail
significant deviations from infinite-$V$ phase behaviors for a 
three-dimensional droplet of linear dimension
$\gtrsim (10^7\times 0.0694/0.7)^{1/3}b\approx 40$ nm.
Notably, this length scale for diminished thermal impact is not dissimilar
to the $\sim 80$ nm estimated above for diminished impact of interfacial 
effects (Fig.~6) to enable effective noise buffering by ternary LLPS.
\\


\noindent
{\large\bf Discussion}\\

Amongst tremendous recent advances, many facets of
multicomponent LLPS \cite{KnowlesPNAS2024,Loew2021,KnowlesPRX2026}
remain unexplored as we seek insights into biologically 
functional condensates which are almost invariably composed of 
multiple biomolecular components. 
In view of the size diversity of biomolecular condensates including
ones that operate at nanoscale \cite{Nano2023}, here we address 
the equilibrium aspects \cite{Safran2021} of the impact of
condensate size and the number of phases in multicomponent 
LLPS on its ability to buffer noise in biomolecular concentrations, 
while deferring considerations of diffusion, phase exchange, and 
other dynamic aspects of concentration buffering and noise 
reduction \cite{Zechner2020,Julicher2025,dynamics} to future studies. 
Beside focusing on the less-explored interplay of finite size and phase 
multiplicity, our effort is also complementary to studies of
concentration buffering and noise reduction in the dilute 
phase \cite{Zechner2020,Julicher2025} wherein the role of condensed phase as 
a sink of over-abundant dilute-phase biomolecules is emphasized. By comparison,
we are more interested in various LLPS processes' 
abilities to maintain constant compositions in multiphasic condensed phases as 
model local microenvironments \cite{Arosio2024,JulieRNA2022}
serving as natural or designed biochemical 
``reaction crucibles'' \cite{crucibles}.

Taking a first step toward elucidating effects of phase multiplicity,
we compare buffering performance in binary and ternary LLPS of pairs of 
polyampholyte sequences \cite{njp2017} (Figs.~1, 2 and S6).  Ternary LLPS bears 
directly on biomolecular condensates with a core-shell architecture 
such as the nucleolus \cite{feric2016} that entail at least a three-phase 
coexistence: a condensed core phase, a condensed shell phase, and a dilute 
phase. As noted, despite being often overlooked, ternary LLPS should in 
principle outperform 
binary LLPS in noise buffering because the ternary coexisting phases are 
invariant over a 2D area instead of only along a 1D binary tieline (Fig.~1). 
Indeed, for natural cell-core condensates such as the nucleolus, it is 
biophysically appealing to surmise that ternary LLPS may first establish
concentrations of the two-phase droplets, to be followed by more complex 
assembly processes \cite{Kriwacki2025} without needing a substantial post-LLPS 
concentration readjustment. Within this general picture,
a major finding of ours is that the effectiveness of 
concentration noise buffering by ternary LLPS-like behaviors is strongly
affected by system size. For MD-simulated droplets of typical linear 
dimension $\sim 15$ -- $50$ nm, our extensive MD data showcase a high 
degree of concentration dispersion, which is 
at variance with conventional infinite-$V$ 
ternary LLPS, resulting in poor noise buffering (Figs.~4, 5 and S6).
Semi-quantitative comparisons of the MD results against those from our 
finite-$V$ phase-field theory (Figs.~3, 5, and 6) and Gibbs-ensemble-like MC 
model (Fig.~7) lead us to conclude that the significant MD concentration 
dispersions are finite-$V$ interfacial and thermal effects, which will
become negligible as $V\rightarrow\infty$. Our analysis concludes further 
that for droplet linear dimension $L\gtrsim 80$ nm, 
our model polyampholyte ternary LLPS-like systems should behave very 
similiarly to their corresponding infinite-$V$ ternary systems and 
thus surpass binary LLPS as a mechanism for concentration noise buffering 
(Fig.~6).
While this estimate may suggest that micron-size biomolecular condensates
and even nanocondensates with $L\lesssim 500$ nm deviate little from
infinite-$V$ LLPS systems, it should be recognized, as mentioned earlier, that 
such an estimate is inherently sensitive to sequences (of proteins and/or 
nucleic acids), chain lengths, and other properties of the biomolecules 
involved, and thus requires further investigations for different systems.
To our knowledge, invariance of three coexisting phases within a 2D
triangular regime on a 2D concentration plane has not been experimentally
tested in a biomolecular context. An earlier experimental measurement of
partitioning of the protein NPM1 into the nucleolus indicated behaviors
more akin to a binary rather than a ternary LLPS \cite{Riback_etal2020}.
It would be instructive if high-throughput 
approaches \cite{KnowlesNatComm2022} can be used to perform experimental 
scans of initial concentrations similar to the MD scan in Fig.~S6
to gain comprehensive information into three-phase behaviors.

In line with a recent application of RPA to predict sequence-dependent 
partitioning of intrinsically disordered proteins into biomolecular 
condensates \cite{BenS2025}, our study underscores analytical 
theory \cite{Analytic,MiMB2023} as an efficient approach to sequence-dependent 
multicomponent LLPS and thus can potentially be utilized for the 
experimental design of shell-core, multilayered and other multiphasic 
condensates \cite{Good2024,RanaCliff2024,Chilkoti2017,CUHK,Shrinivas2026}. Analytical
theory should be particularly useful in situations when the driving force
for demixing is subtle \cite{njp2017,Pal2021,RanaCliff2024} because there is 
no clear distinction---as in the present case of electrostatics---between 
homotypic and heterotypic interactions. For instance, RPA predicts that the 
2D phase diagrams of certain oppositely charged sequence pairs 
have two distinct ternary 
LLPS regions (Fig.~8a--c). The corresponding phase-field condensed state 
has a core-shell architecture with a mixed core phase whose
composition can be changed 
by varying the initial concentrations (Fig.~8d--f; see additional examples in 
Fig.~S14).  It would be exciting to investigate whether such an intriguing 
feature is realizable experimentally.
Moreover, our application of phase-field theory to MD data to explore 
system size dependence (Figs.~3, 5, and 6) suggests that it may be possible 
to develop similar yet more rigorous analytical approaches to transform 
results from MD-simulated LLPS systems of relative small sizes to 
biologically realistic length scales. In this regard, it has been
stipulated recently that a model system with $\gtrsim 10^4$ molecules
is needed for an accurate estimation of the critical temperature $T_{\rm cr}$
of LLPS \cite{Rohit2026}. This system size threshold
is within an order of magnitude of the above $L\gtrsim 200b\approx 80$ nm 
condition for diminished finite-$V$ effects in our model droplets because
the number of 50-bead chains in such a droplet with $\rho \approx 0.7b^{-3}$ 
(Fig.~3a) is $\approx (200b)^3\rho/50\approx 10^5$. Indeed, the similarity 
of these two size estimates is not unexpected because ideal $T_{\rm cr}$ and
ternary LLPS properties both requires $V\rightarrow\infty$.
All in all, the conceptual framework developed above from simple model 
sequences regarding the impact of droplet size and phase multiplicity on 
concentration noise buffering should contribute to future investigations
of this important feature in biomolecular condensate in settings with 
higher degrees of biological realism.
\\

\noindent
{\large\bf Materials and Methods}\\

{\bf RPA.}
Following our previous developments of RPA theory for 
sequence-dependent LLPS \cite{Analytic,MiMB2023},
the free energy density/$k_{\rm B}T$ for a multicomponent system containing
$n_p$ copies of individual polymer species (labeled by $p$) of chain 
length $N_p$ carrying net charge $q_p$ and $n_{\rm w}$ solvent molecules 
with total (w + polymers) density $\rho_{\rm tot}$ is given by 
\begin{equation}
\label{eq:rpa_free_energy}
\begin{aligned} 
f_{\rm RPA} & = \phi_{\rm w} \ln \phi_{\rm w} + 
\sum_p \frac{\phi_p}{N_p} \ln \phi_p + 
\frac{2 \pi \lB \rho_{\rm tot} }{\kappa_{\rm D}^2} \left( \sum_p 
\frac{q_p}{N_p} \phi_p \right)^2 \\
& 
\quad\quad\quad\quad\quad\quad\quad\quad
+ \frac{1}{\rho_{\rm tot}} \int \frac{\dd \kk }{(2 \pi)^3} 
\ln\left(1 + \frac{4 \pi \lB \rho_{\rm tot}}{\kappa_{\rm D}^2 + \kk^2} 
\sum_p \phi_p g_p(|\kk|) \right) \; .
\end{aligned}
\end{equation}
Here $\phi_p = N_p n_p / V \rho_{\rm tot}$, 
$\phi_{\rm w} = n_{\rm w}/V \rho_{\rm tot} = 1 - \sum_p \phi_p$ 
are volume fractions, where incompressibility is assumed by the latter
equality, $\kappa_{\rm D}$ is inverse Debye screening length, $\kk$ is
vectorial position in reciprocal space, and sequence dependence is 
approximately accounted for by
$g_p(|\kk|) = \sum_{\mu=1}^{N_p} \sum_{\nu=1}^{N_p} \sigma_{p,\mu} 
\sigma_{p,\nu} \exp(-|\mu - \nu| |\kk|^2 b^2 / 6 - a_{\rm s}^2 |\kk|^2)/N_p$
wherein $\sigma_{p,\mu}$ is the charge of the $\mu$th bead along polymer 
$p$ (thus $q_p = \sum_{\mu=1}^{N_p} \sigma_{p,\mu}$), and $a_{\rm s}$
is the standard deviation of the Gaussian smearing for regularizing
short-distance divergences \cite{MiMB2023}.
Derivation of Eq.~\ref{eq:rpa_free_energy} from the full coordinate-space 
Hamiltonian and other details of our formal approach are provided in SI Text.
\\

{\bf Coarse-grained MD.}
Simulations are conducted by the ``Slab'' approach \cite{dignon18}
at the $T^*$s indicated above using our polyampholyte model in 
refs.~\citen{SumanPCCP2018,Pal2021} with the same harmonic bonded term and non-bonded 
electrostatic interactions as well as a ``1/3 LJ'' potential to provide
excluded volume---but with a reduced sequence-nonspecific overall attractions 
to facilitate comparison with RPA predictions derived in the absence of
such attractions \cite{SumanPCCP2018,Pal2021}. 
The dimensions of the simulation box and the Langevin dynamics parameters
are also identical to those in refs.~\citen{SumanPCCP2018,Pal2021}.
All MD-simulated systems in this work contain a total of 500 chains 
except for the 1,000-chain study in Fig.~S9.

{\bf Phase-field theory.}
The equilibrium spatial density distributions 
governed by Eq.~\ref{F_eq} are approximated 
by free energy minimization (without considering thermal fluctuations): 
\begin{equation} 
\frac{\delta F}{\delta \rho_p(\rr)} = -\sum_q \xi_{p,q} 
\nabla^2 \rho_q(\rr) + 
\frac{\partial f_{\rm RPA}}{\partial \rho_p}(\lbrace \rho_q(\rr) \rbrace) 
= 0
\end{equation}
constrained by the conservation of bead number, i.e., 
$\int \dd \rr \rho_p(\rr) / V = \bar{\rho}_p$.
Details of the formulation are provided in SI Text.

{\bf MC moves.}
Moves are designed to satisfy both the Boltzmann exponents and the 
$\delta$-function constraints on $\phi$s in Eq.~\ref{Z_eq}. The sampling
protocol for the FH model in Fig.~7a--d is as follows:
(i) Each voxel is initialized with an input $\bar{\phi}$ equals to one of 
the discretized $\phi$ values.
(ii) Two voxels $k,k^\prime$ with existing
$\phi=\phi^{(k)}_{\mathrm{old}},\phi^{(k^\prime)}_{\mathrm{old}}$ 
are randomly selected. 
(iii) A new $\phi_{\mathrm{new}}$ is randomly selected among the 
discretized $\phi$ values to replace
$\phi^{(k)}_{\mathrm{old}}$.
(iv) $\phi^{(k^\prime)}_{\mathrm{old}}$ is then replaced by
$\phi^{(k^\prime)}_{\mathrm{new}}=[\phi^{(k^\prime)}_{\mathrm{old}}-
(\phi^{(k)}_{\mathrm{new}} - \phi^{(k)}_{\mathrm{old}})]$
such that $\bar{\phi}$ is not changed.
(v) This attempted move is rejected if 
$\phi^{(k)}_{\mathrm{new}}$ or $\phi^{(k^\prime)}_{\mathrm{new}}$ is
$<0$ or $>1$.
(vi) Let $\Delta f=
f(\phi^{(k)}_{\mathrm{new}})+
f(\phi^{(k^\prime)}_{\mathrm{new}})-
f(\phi^{(k)}_{\mathrm{old}})-
f(\phi^{(k^\prime)}_{\mathrm{old}})$.
The attempted move is accepted if a random number $\in [0,1]$ is 
$\leq e^{-V_v \Delta f}$.
Equilibration criterion and the corresponding moves for two-polymer
systems (Fig.~7e--g and Fig.~S13) are provided in SI Text.
\\

$\null$\\
{\large\bf Acknowledgments}\\ 
We thank Julie Forman-Kay for helpful discussions.
This work was supported by Canadian Institutes of Health Research (CIHR) grant
NJT-155930, Natural Sciences and Engineering Research Council of Canada (NSERC)
grants RGPIN-2018-04351 and RGPIN-2024-04167, and the computational resources
provided by the Digital Research Alliance of Canada.

\clearpage

\setcounter{figure}{0}
\renewcommand{\figurename}{{\bf Fig.}}
\renewcommand{\thefigure}{{\bf \arabic{figure}}}

\begin{figure}[t]
   \centering
   \includegraphics[width=\columnwidth]{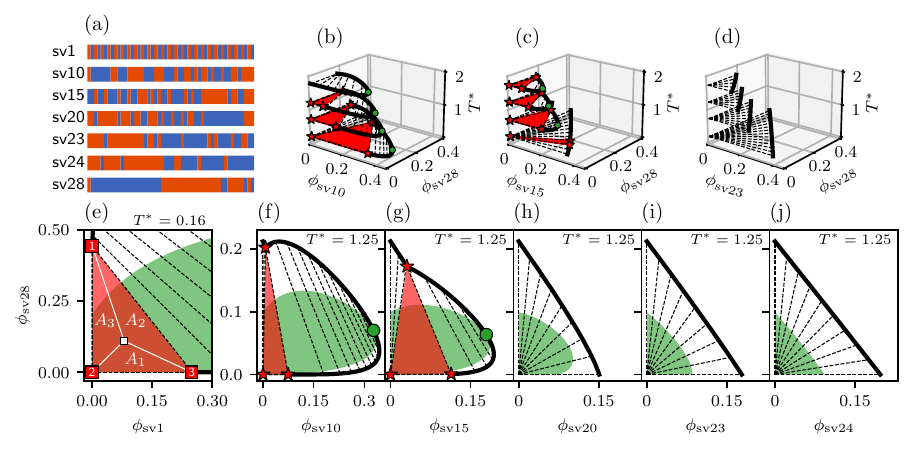}
   \caption{Sequence- and $T$-dependent ternary and binary 
LLPS. (a) Model 50-bead ``sv" sequences (as labeled) studied, 
where positive (blue) and negative (red) bead positions are marked by stripes. 
(b)--(d) RPA-predicted phase diagrams of pairs of sv sequences
indicated by the subscripts of volume fraction $\phi$; 
$T^*=b/l_{\rm B}\propto T$ is 
reduced temperature (see text). Each red triangle is a ternary LLPS regime 
with three coexisting phases at the vertices (red stars).
Dashed lines are tielines connecting two coexisting phases 
on the binodal boundary (solid black curve) for binary LLPS.
(e) Phase diagram at
$T^*= 0.16$ (Bjerrum length $l_{\rm B}=6.25b$, inverse Debye screening length
$\kappa_{\rm D}=0$) showing the spinodal regime 
(green) and binary tielines (dashed). 
As an illustration for a ternary regime (translucent red triangle),
the three coexisting phases for any 
$(\bar{\phi}_{\rm sv1},\bar{\phi}_{\rm sv28})$
of overall volume fractions (white square) 
are numbered 1, 2, 3 (red squares) with
fractional populations proportional to the triangular areas
$A_1$, $A_2$, $A_3$ (ref.~\citen{njp2017}).
(f)--(j) Phase diagrams for different sequence pairs 
(all at $T^*=1.25$, $l_{\rm B}=0.8b$, $\kappa_{\rm D}=0$) 
showing---where applicable---ternary 
[translucent red, as in (e)] and spinodal (green) regimes,
binodal boundary (solid black), binary tielines (dashed), 
ternary coexisting phases (red stars),
and critical points (green circles).
}
   \label{fig1}
\end{figure}
\vfill\eject

\begin{figure}[t]
   \centering
   \includegraphics[width=\columnwidth]{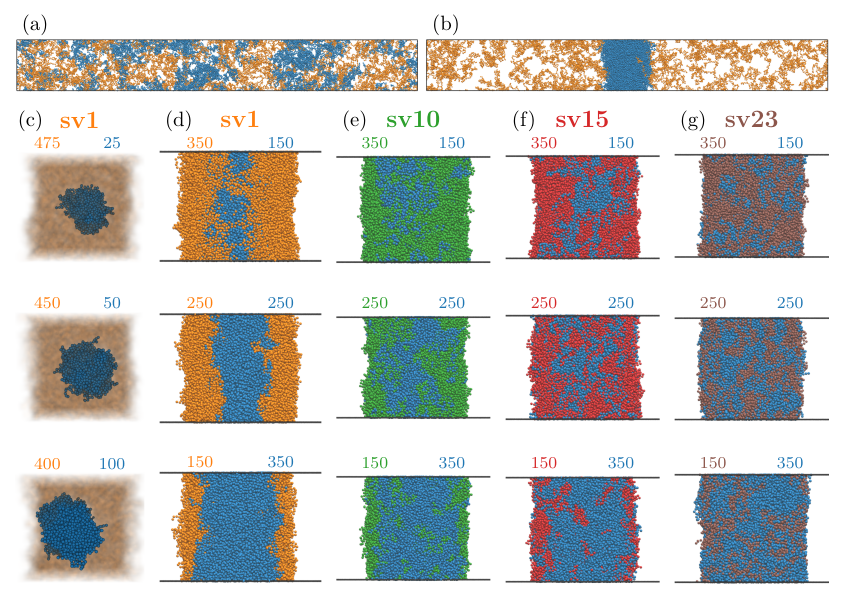}
   \caption{Explicit-chain MD snapshots 
of binary and ternary LLPS-like behaviors. Each simulation was for
a total of 500 chains with periodic boundary conditions. (a,b) Mixtures 
of sv1 (golden) and sv28 (blue)
at (a) $T^*=4.0$ with no LLPS and (b) $T^*=1.4$ exhibiting 
binary but not ternary LLPS. Images 
in (a,b) are adapted from our previous study \cite{Pal2021}.
(c--g) Condensed droplets of different sequences (color coded)
pairing with sv28 (blue) in different concentration/volume fraction 
proportions (all at $T^*=0.65$) indicated by the numbers 
of simulated chains for each sequence using the same color code.
(c) A translucent rendering of 
sv1 density for better depiction of the sv28-rich droplet with 
an explicit representation of the sv28 beads.
(d--g) Chain beads of both sequences are represented explicitly.
Ternary LLPS-like features are showcased in (c,d).
}
   \label{fig2}
\end{figure}
\vfill\eject

\begin{figure}[t]
   \centering
   \includegraphics[width=\columnwidth]{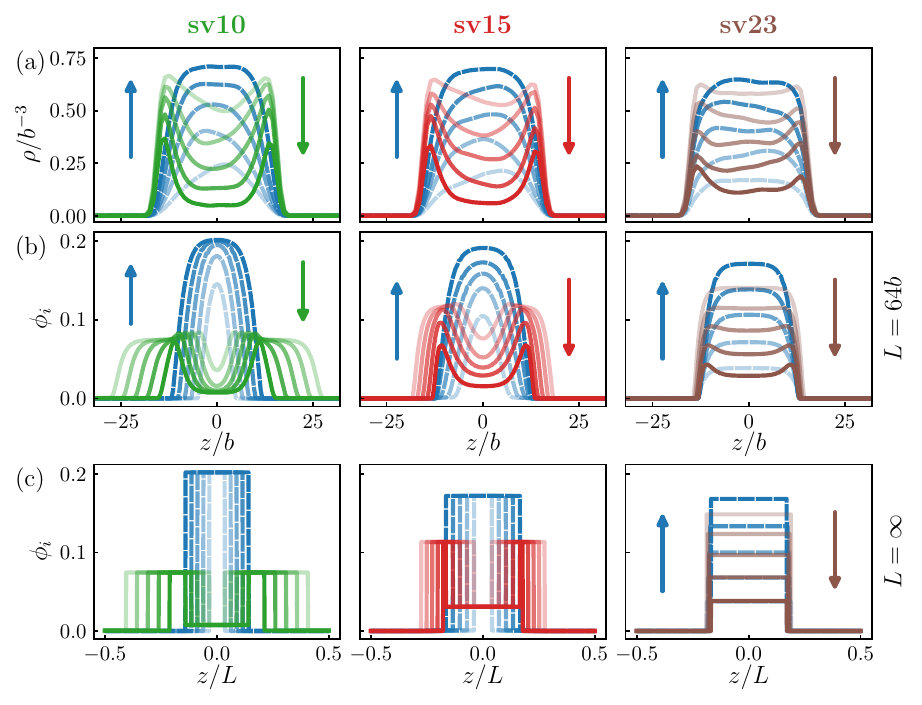}
   \caption{Phase-field theory captures 
essential variation in 1D MD density profiles. 
Color-coded profiles for increasing 
(up arrow) or decreasing (down arrow) initial (overall) 
$\bar{\rho}$ or $\bar{\phi}$ of sequences are shown in different shades.
(a) $\rho(z)$s from Fig.~S6 
for number of chains $(n_{i},n_{\rm sv 28}) 
= (400,100)$, $(325,175)$, ($250,250)$, $(175,325)$ 
and $(100,400)$, where $i=$ sv10, sv15, and sv28.
(b) Corresponding $\phi(z)$s computed numerically using 1D 
phase-field theory at $l_{\rm B} = 0.8b$, $\kappa_{\rm D}=0$, and $L=64b$ 
with $\bar{\phi}_{i} + \bar{\phi}_{\rm sv28} = 0.07$
by discretizing $z$
into $N_z=150$ evenly spaced lattice points with periodic boundary conditions 
for $(\bar{\phi}_i,\bar{\phi}_{\rm sv28})=(0.057,0.013)$, 
$(0.046,0.024)$, $(0.036,0.034)$, $(0.025,0.045)$ and $(0.014,0.056)$.
(c) Theoretical $\phi(z)$s in the 
infinite-volume $L\rightarrow\infty$ limit. 
}
   \label{fig4}
\end{figure}
\vfill\eject

\begin{figure}[t]
   \centering
   \includegraphics[width=\columnwidth]{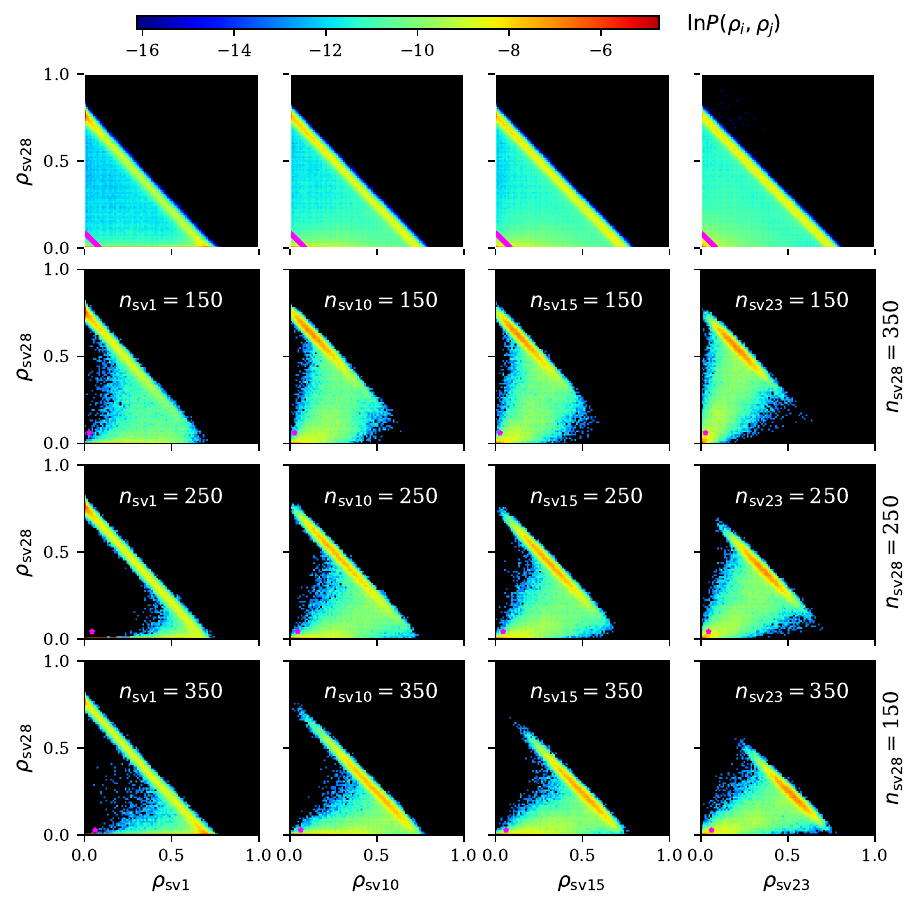}
   \caption{MD-simulated normalized equilibrium distribution
$P(\rho_i,\rho_j)$ of local bead 
densities $\rho_i,\rho_j$ for a sequence pair. The logarithmic heat maps 
(top scale, $P=0$ depicted as black) are for $i$-$j=$
sv1-sv28, sv10-sv28, sv15-sv28, and sv23-sv28 
at $T^*=0.65$ (same data as that for Fig.~S6).
The top-row shows equal-weighted averages, $\langle P(\rho_i,\rho_j)\rangle$s, 
over all initial input $(\bar{\rho}_i,\bar{\rho}_j)$ corresponding to 
$(n_i,n_j)=$ $(25, 475), (50, 450),\dots, (475, 25)$ in Fig.~S6. 
The lower three rows are for the $(n_i,n_j)$s indicated.
Local densities are determined from 10,000 MD snapshots as numbers of beads 
of a given sequence found in $6\times 6\times 6b^3$ cells (voxels) 
within the simulation box divided by the voxel volume $6^3$ in units of $b^3$.
$P(\rho_i,\rho_j)$ is then constructed by binning these local densities
into a $100\times 100$ grid of $\rho_i,\rho_j \in [0,1]$. 
Distributions obtained using an alternate $8\times 8\times 8b^3$ voxel 
are very similar.
Initial densities marked by the purple stars are given by 
$(\bar{\rho}_i,\bar{\rho}_j)$ $=$ $(n_i/V_{\rm MD},n_j/V_{\rm MD})$ 
where $V_{\rm MD}=33\times 33\times 264b^3$ is the total volume of 
the MD simulation box.
}
   \label{fig5}
\end{figure}
\vfill\eject

\begin{figure}[t]
   \centering
   \includegraphics[width=0.80\columnwidth]{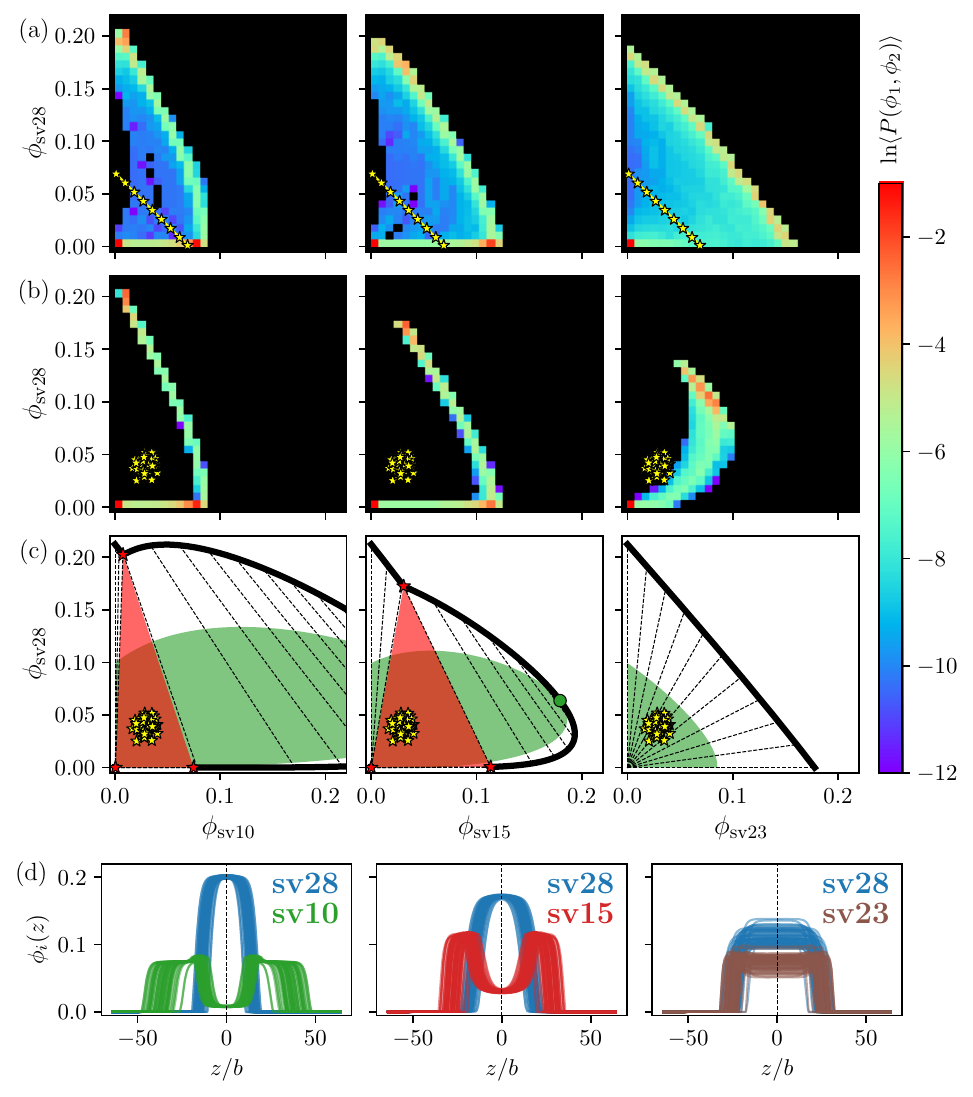}
\vskip -0.4cm
   \caption{Phase-field model of finite-size concentration distributions of 
sequence pairs. Results are computed using
$l_{\rm B}=0.8b$, $L=128b$ and $N_z = 150$ in the 1D phase-field theory. 
(a,b) The heat maps show phase-field model $\langle P(\phi_i,\phi_j)\rangle$ 
distributions for three sequence pairs, each averaged over the 
initial $(\bar{\phi}_i,\bar{\phi}_2)$ (yellow stars) with 
equal weights. The logarithmic scale for $P>0$ is shown on the right;
$P=0$ is depicted in black.  
To better capture the rapid variations in $\phi$ at phase 
boundaries, the 1D phase-field $\phi$ profiles (exemplified
by Fig.~3b) are refined by iteratively interpolating linearly
between every pair of neighboring lattice points for 10 iterations.
The $P(\phi_i,\phi_j)$s represented by the heat maps are then constructed 
by binning the resulting $(\phi_i, \phi_{\rm sv28})$ 
at the original $N_z = 150$ and interpolated 1D lattice points
onto a $30\times 30$ grid 
with $\phi_{i,{\rm sv28}} \in [0,0.22]$, where $i=$ sv10, sv15, or sv28.
The 64 input $(\bar{\phi}_i,\bar{\phi}_{\rm sv28})$
are (a) evenly spaced on the 
$\phi_i + \phi_{\rm sv28} = 0.07$ line or (b) drawn from normal distributions 
with means $\langle \bar{\phi}_i \rangle = 0.03$, 
$\langle \bar{\phi}_{\rm sv28} \rangle = 0.04$, and standard deviations
$\sigma_{\bar{\phi}_i} = \sigma_{\bar{\phi}_{\rm sv28}} = 0.005$. 
(c) Distribution of input $\bar{\phi}$s
superposed on the corresponding RPA phase diagrams 
for infinite $V$ (same notation as that in Fig.~1f--j).
(d) The phase-field model $\phi$ profiles (color coded for
sequence pairs) for the multiple input $\bar{\phi}$s in (b,c). 
}
   \label{fig6}
\end{figure}
\vfill\eject

\begin{figure}[t]
   \centering
   \includegraphics[width=0.75\columnwidth]{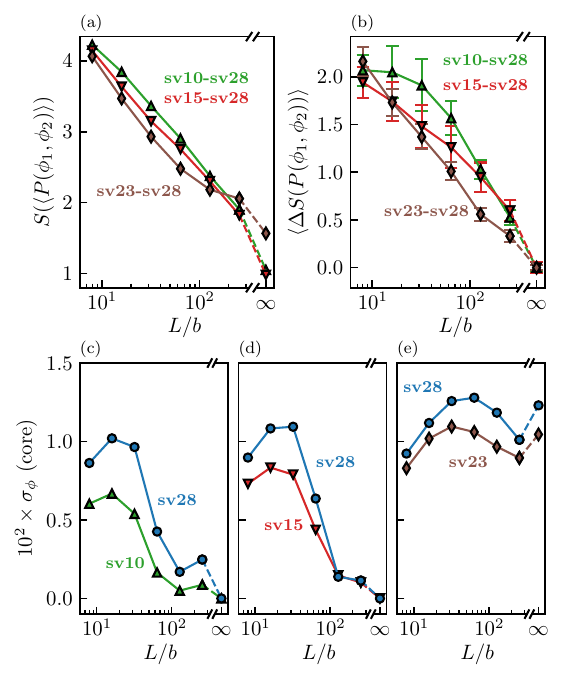}
   \caption{Noise buffering by ternary and binary LLPS depends on system
size. Analysis is based on 1D phase-field $\phi$ profiles 
at $l_{\rm B} = 0.8b$, $\kappa_{\rm D}=0$,
system sizes $L/b = 8$, $16$, $32$, $64$, $256$, and in the
$L\rightarrow\infty$ limit for the set of normal-distributed
input $\{(\bar{\phi}_i,\bar{\phi}_{\rm sv28})\}$ in Fig.~5b,c.
(a) $S(\langle P(\phi_1,\phi_2)\rangle)$ quantifies noise buffering for the
set of multiple inputs.
(b) $\langle \Delta S(P(\phi_1,\phi_2))\rangle=
\langle[S(P(\phi_1,\phi_2))-S(P(\phi_1,\phi_2);L=\infty)]\rangle$
shows average of $S$ for individual inputs
as it approaches its infinite-$L$, 
$\Delta S(P)=0$ limit. 
(c--e) The standard deviation $\sigma_\phi$ of $\phi(z=0)$
in the condensed core.
}
   \label{fig7}
\end{figure}
\vfill\eject

\begin{figure}[t]
   \centering
   \includegraphics[width=\columnwidth]{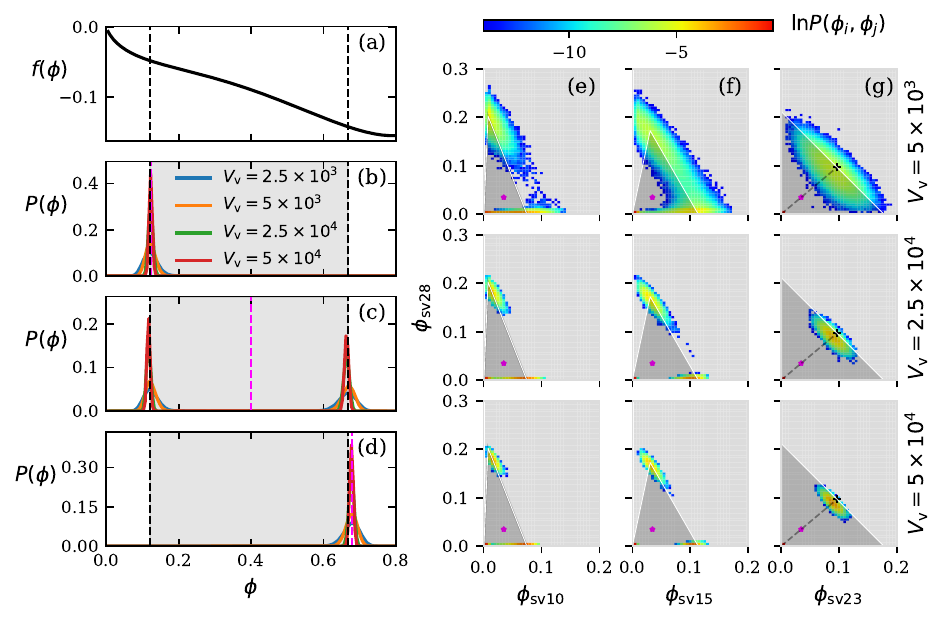}
   \caption{Finite-size Gibbs-ensemble-like MC formulation partially 
rationalizes diffused MD-simulated density distributions.
Results are for $V=200V_{\rm v}$. 
(a--d) One-solute FH phase behaviors for different $V_{\rm v}$s in FH 
lattice units.
(a) Free energy/$Vk_{\rm B}T$ as a 
function of solute volume fraction $\phi$ for a FH model with
$N=3$ monomers per chain and $\chi=$ $1.4117$;
$\phi_{\rm dil}$ and $\phi_{\rm con}$ are marked by dashed lines.
(b--d) Equilibrium distribution of $\phi$ for
different $V_{\rm v}$s [color code in (b)] with initial (input) 
$\bar{\phi}=$ $0.1238$ (b), $0.4010$ (c), or $0.6782$ (d)
(purple dashed line). The LLPS region is shaded gray.
(e--g) $V_{\rm v}$-dependent equilibrium $\phi$ distributions based 
on $f_{\rm RPA}$ [Eq.~6] with $V_{\rm v}$ in units of $b^3$ for 
(e) sv10--sv28, (f) sv15--sv28, and
(g) sv23--sv28, all computed at $l_{\rm B}=0.8b$ and $\kappa_{\rm D}=0$
for input $(\bar{\phi}_i, \bar{\phi}_{\rm sv28})= 
(0.0347, 0.0347)$ indicated by the purple star. 
The heat maps show $P>0$ distributions (top logarithmic scale)
whereas $P=0$ is in light gray. The dark gray triangles 
are the RPA-predicted ternary (e,f) or binary (g) LLPS areas; the dashed
tieline (g) is for $(\bar{\phi}_{\rm sv23}, \bar{\phi}_{\rm sv28})$.
}
   \label{fig8}
\end{figure}
\vfill\eject

\begin{figure}[t]
   \centering
   \includegraphics[width=0.8\columnwidth]{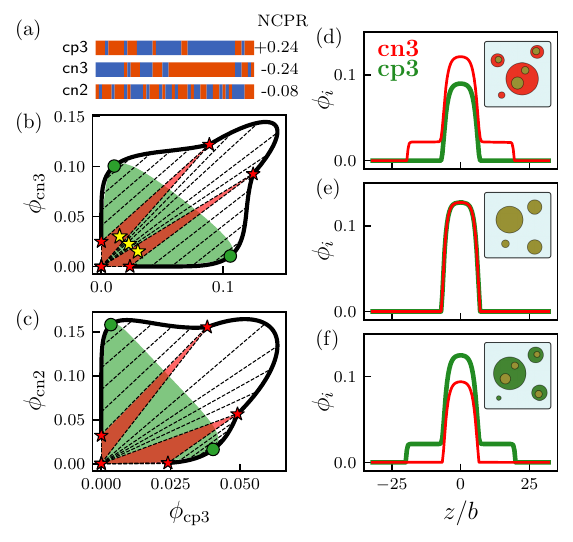}
   \caption{Examples of charge sequence pairs with multiple ternary 
LLPS regimes. (a) 50mer charged sequences with different
net charge per residue (NCPR) values 
in the style of Fig.~1a. 
(b,c) RPA-predicted phase diagrams for the (b) cn3-cp3 
and (c) cn2-cp3 pairs at $l_{\rm B}=2b$ and 
$\kappa_{\rm D}=0.5b^{-1}$ in the style of Fig.~1f-j.
(d) Phase field theory volume fraction profiles (color coded
curves) for the same $l_{\rm B},\kappa_{\rm D}$ with 
system size $L=64b$, $N_z=150$, and 
input $(\bar{\phi}_{\rm cp3},\bar{\phi}_{\rm cn3})=$ $(0.015,0.03)$ (d),
$(0.0225,0.0225)$ (e), and
$(0.03, 0.015)$ (f) as marked by the yellow stars in (b).
The different phase behaviors in (d--f) are illustrated by the
inset cartoons using the same color code, wherein mixtures of
different proportions of cn3 and cp3 are represented by different shades
of red and green.
}
   \label{fig9}
\end{figure}
\vfill\eject


\vfill\eject
\clearpage
\DeclareRobustCommand{\citen}[1]{%
  \begingroup
    \romannumeral-`\x 
    \setcitestyle{numbers}%
    \cite{#1}%
  \endgroup
}
%

\setcounter{equation}{0}
\renewcommand{\theequation}{{\rm S}\arabic{equation}}
%
%
%

$\null$
\hfill {\bf July 12, 2026}
\vskip 0.3in

\begin{center}

{\Huge\bf Supporting Information}\\

\vskip 0.3cm

{\Huge ({\it SI Appendix})}\\

\vskip 0.3cm

{\Large\it for}

\vskip 0.3cm

{\Large\bf Homeostatic Noise Buffering in Biomolecular}\\

\vskip 0.3cm

{\Large\bf Condensates Hinges on Phase Multiplicity}\\

\vskip 0.3cm

{\Large\bf Modulated by Interfacial and Droplet Size Effects}\\

\vskip .5in
{\bf Jonas W{\footnotesize{\bf{ESS\'EN}}}},$^{1,\dagger}$
{\bf Tanmoy P{\footnotesize{\bf{AL}}}},$^{1,\dagger}$
{\bf Suman D{\footnotesize{\bf{AS}}}},$^{2,\dagger}$
and
{\bf Hue Sun C{\footnotesize{\bf{HAN}}}}$^{1,*}$

$\null$

$^1$Department of Biochemistry,
University of Toronto, Toronto, Ontario M5S 1A8, Canada\\
$^2$Department of Chemistry, Gandhi Institute of Technology and
Management, Visakhapatnam, Andhra Pradesh 530045, India\\

\vskip 1.3cm

%

\end{center}

$\null$\\

\noindent
$^\dagger$Contributed equally.

\vskip 1.3cm

\noindent
$^*$Correspondence information:\\
{\phantom{$^\dagger$}}
Hue Sun C{\footnotesize{HAN}}.$\quad$
E-mail: {\tt huesun.chan@utoronto.ca}\\
{\phantom{$^\dagger$}}
Tel: (416)978-2697; Fax: (416)978-8548\\
{\phantom{$^\dagger$}}
Department of Biochemistry, University of Toronto,
Medical Sciences Building -- 5th Fl.,\\
{\phantom{$^\dagger$}}
1 King's College Circle, Toronto, Ontario M5S 1A8, Canada.\\

\vfill\eject

\noindent{\Large\bf Supporting Materials and Methods}
\\

This document provides details of the analytical formulations
and associated numerical protocols developed for this work. Also
included are supporting Table~S1 and supporting figures
 Figs.~S1--S14 shown in the order by which they are referred to
in the maintext.
\\

\noindent
{\large\bf Analytical Theories of Sequence-Dependent Phase Separation}
\\

\noindent
{\bf Model definition and the field-theoretic framework}

We consider a system of volume $V$ containing $n_{\rm w}$ solvent molecules
(``w'' stands for water in biological systems but the formulation is general
and not restricted to a particular solvent)
and $M$ chemically distinct components of bead-spring polymers. We let $N_p$
represent the number of beads on a chain of type-$p$ (with $p=1,\dots,M$) and
denote the charge sequence of type-$p$ polymers as $\sigma_{p,\alpha}$ (with
$\alpha=1,\dots,N_p$), i.e.~$\sigma_{p,\alpha}$ is the electric charge of bead
$\alpha$ on any of the type-$p$ chains (``$\alpha$'' here should not be 
confused with the phase label that we have used in another context). All 
charges are given in units of the proton charge denoted here as $e$; and $n_p$
is the number of $p$-type chains in the system.
The position of bead $\alpha$ on the $i$th chain of the $p$th polymeric
component is denoted as $\RR_{p,i,\alpha}$ (with $i=1,\dots,n_p$) and the
locations of the solvent molecules are denoted as $\rr_i$ (with
$i=1,\dots,n_{\rm w}$).

The model is defined by the Hamiltonian
\begin{equation} \label{eq:particle_hamiltonian}
\begin{aligned}
\hat{H} = & \frac{3}{2b^2}\sum_{p=1}^M \sum_{i=1}^{n_p} 
\sum_{\alpha=1}^{N_p -1} 
(\RR_{p,i,\alpha+1}-\RR_{p,i,\alpha})^2 
               +\frac{1}{2\gamma_{\rm ex}} \int \dd \rr 
\left( \sum_{p=1}^{M} \hat{\rho}_p(\rr) + \hat{\rho}_{\rm w}(\rr) \right)^2 
\\
& \hspace{3cm}
               + \frac{\lB}{2}\int \dd \rr \int \dd \rr' \, \frac{\hat{c}(\rr) \hat{c}(\rr')}{|\rr-\rr'|}\e^{-\kappa_{\rm D} |\rr - \rr'|}
\end{aligned}
\end{equation}
where the three terms account for chain-connectivity, contact excluded-volume
repulsion, and electrostatic interactions, respectively. Chain-connectivity is
implemented through spring-like forces between consecutive beads, with a common
Kuhn length $b$ for all bonds and polymer species.

The strength of the short-range repulsive excluded-volume interactions, 
provided by the second term in Eq.~\eqref{eq:particle_hamiltonian}, is set by 
$\gamma_{\rm ex}^{-1}$ where $\gamma_{\rm ex}$ has units of 
$(\mbox{volume})^{-1}$ and characterizes the magnitude of spatial fluctuations
in total density. This term is physically equivalent to a
``soft-compressibility'' term previously considered\cite{WessenJCP} and is
formulated in terms of the microscopic polymer bead- and solvent number
densities $\hat{\rho}_{p}(\rr)$ and $\hat{\rho}_{\rm w}(\rr)$, respectively,
given by 
\begin{equation} \label{eq:micro_density}
\hat{\rho}_{p}(\rr) = \sum_{i=1}^{n_p} 
\sum_{\alpha=1}^{N_p} \Gamma(\rr-\RR_{p,i,\alpha}) \; , 
\quad\quad
\hat{\rho}_{\rm w}(\rr) = \sum_{i=1}^{n_{\rm w}} \Gamma(\rr-\rr_i) \; .
\end{equation}
Here, $\Gamma(\rr) =(1/2\pi a_{\rm s}^2)^{3/2} \e^{-\rr^2/2a_{\rm s}^2}$
serves to regularize infinities arising from self- and contact interactions by
``smearing'' every particle over a Gaussian distribution with a small width
$a_{\rm s}$ (ref.~\citen{Wang2010}). 
For simplicity, we assign a common 
smearing length $a_{\rm s}=b/\sqrt{6}$ for all particles in all
subsequent numerical computations. 

The third term in Eq.~\eqref{eq:particle_hamiltonian}, describing
electrostatic interactions, is expressed in terms of the microscopic charge
density $\hat{c}(\rr)$ given by 
\begin{equation}
\label{eq:micro_charge_density}
\hat{c}(\rr) = \sum_{p=1}^M \sum_{i=1}^{n_p} 
\sum_{\alpha=1}^{N_p} \sigma_{p,\alpha} \Gamma(\rr - \RR_{p,i,\alpha}).
\end{equation}
The Bjerrum length $\lB = e^2/4 \pi \epsilon k_{\rm B} T$ controls the
strength of electrostatic interactions, while salt is included implicitly
through a Debye screening length $\kappa_{\rm D}^{-1} 
= \sqrt{4 \pi \lB I} $ where $I$
is the solution ionic strength, e.g.~$I=2[{\rm NaCl}]$ in the case of (fully
dissociated) NaCl. Our choice of considering implicit salt through Debye
screening, rather than explicit salt ions, simplifies phase diagram
calculations since explicit salt ions would imply an additional chemical
potential that would have to be included in phase co-existence conditions.
Although the implicit ion picture imposes a serious approximation since
effects from ion partitioning are ignored, we expect all charge-pattern trends
observed in this work to be robust with respect to explicit/implicit salt. For
situations where salt-partitioning is expected to be crucial, it is
straight-forward to extend field theories, like the one considered in this
work, to include explicit ions\cite{WessenJCP,Analytic}.
As a measure of interaction strength, the reduced temperature $T^*$ is 
defined as the inverse Bjerrum length in units of $b$ in our formulation, 
i.e., 
\begin{equation}
T^* = \frac{b}{\lB} \; .
\end{equation}
The statistical mechanical partition function of the system described by
$\hat{H}$ can be converted into that of a statistical field theory using
mathematical techniques that can, by now, be considered
standard\cite{Edwards1965,Fredrickson2002,Fredrikson2023_book,MiMB2023}. The
statistical field theory is obtained by introducing two fields, $\eta(\rr)$
and $\psi(\rr)$, that are conjugate to the microscopic mass- and charge
densities $\sum_p \hat{\rho}_p(\rr) + \hat{\rho}_{\rm w}(\rr)$ and
$\hat{c}(\rr)$, respectively. This calculation, which is shown in another
section below under the heading ``The field-theory representation of 
multiple-chain polymer systems'', results 
in the statistical field theory partition function 
\begin{equation} \label{eq:field_partition_function}
Z = \frac{V^\nw}{\nw!} \prod_{p=1}^M \frac{V^{n_p}}{n_p!} 
\int \DD \eta \int \DD\psi \, \e^{-H[\eta,\psi]} \; ,
\end{equation}
where the field Hamiltonian is 
\begin{equation} \label{eq:field_hamiltonian}
H = - \sum_{i={\rm w},1}^M n_i \ln Q_i + \int \dd \rr 
\left( \frac{\gamma_{\rm ex} \eta^2}{2}  + \psi \frac{\kappa_{\rm D}^2 - 
{\nabla}^2}{8 \pi \lB} \psi  \right).
\end{equation}
The pre-factors in Eq.~\eqref{eq:field_partition_function} account for
translational entropy. The field Hamiltonian contains single-molecule
partition functions $Q_i[\breve{\eta},\breve{\psi}]$ (with $i={\rm
w},1,\dots,M$) that have functional dependence on Gaussian smeared fields
$\breve{\varphi}(\rr) = \Gamma \star \varphi(\rr) = \int \dd \rr'
\Gamma(\rr'-\rr) \varphi(\rr')$, for $\varphi=\eta,\psi$. We give the full
expressions for $Q_i$ in 
the section ``The field-theory representation of 
multiple-chain polymer systems'' below. Here, we only
note that the smearing removes the short wavelength modes of the fields
(i.e.,~resulting in $Q_i[\breve{\eta},\breve{\psi}]$ depending only 
on Fourier components of
the fields with wavelengths $|\kk|^{-1}\gtrsim a_{\rm s}$). Consequently,
fluctuation modes on length scales $\lesssim a_{\rm s}$ of $\eta(\rr)$ and
$\psi(\rr)$ are only present in the quadratic terms in $H$ (i.e.~the terms
under the spatial integral in Eq.~\eqref{eq:field_hamiltonian}) and therefore
can be
factorized out from $Z$. The contribution to $Z$ from fluctuations on small
length scales are thus independent of the molecule numbers $n_{{\rm w},
1,\dots,M}$ and can therefore be ``renormalized away'' from any physical
observable e.g.~by using the empty system ($n_{\rm w}=n_1=\dots=n_M=0$)
as a reference state, rendering the theory free from ultra-violet (UV)
divergences.
\\

\noindent{\bf The analytical Flory-Huggins (FH) and random-phase 
approximation (RPA)}

To study the field theory in Eq.~\eqref{eq:field_partition_function}, we start
by approximately solving the functional integrals to the minimal order in
field fluctuations that can capture sequence-dependent phase behavior. 
The procedure is detailed in another section below under the heading 
``Derviation of the free energy density in the approximate analytical
approach'',
here we outline the calculation and state the resulting expression for 
the free energy density.  We first simplify the system by taking the limit 
$\gamma_{\rm ex} \rightarrow 0$
which has the effect of imposing an incompressibility condition
\begin{equation}
\sum_{p=1}^M \hat{\rho}_p(\rr) + \hat{\rho}_{\rm w}(\rr)  
= \rho_{\rm tot} \quad (=\mbox{constant in $\rr$}),
\end{equation}
such that the solvent ceases to be an independent component (i.e.~$
\hat{\rho}_{\rm w}(\rr)$ is given by $\hat{\rho}_{1,\dots,M}(\rr)$ through the
above condition). To see this, note that the energetic cost of a spatially
varying total density $\rho_{\rm tot}=\sum_{p=1}^M \hat{\rho}_p(\rr) 
+ \hat{\rho}_{\rm w}(\rr)$ relative to the spatially homogeneous state 
is $\propto \gamma_{\rm ex}^{-1}$, thus variations of total density
$\rho_{\rm tot}$ vanish as $\gamma_{\rm ex} \rightarrow 0 $.

In the analytical approach considered in this work, we do not consider effects
from the density-conjugate field $\eta(\rr)$ beyond imposing the
incompressibility condition. The functional integral over the charge-conjugate
field $\psi(\rr)$, on the other hand, is solved up to quadratic fluctuations,
an approximation often referred to as the random phase approximation (RPA).
The resulting free energy density, $f \equiv - (V \rho_{\rm tot})^{-1} \ln Z
\approx f_{\rm RPA}$, takes the form of a multi-component Flory-Huggins theory
supplemented by a term accounting for sequence-dependent charge-density
fluctuations, 
\begin{equation} \label{eq:rpa_free_energy}
\begin{aligned}
f_{\rm RPA} = & \phi_{\rm w} \ln \phi_{\rm w} + \sum_{p=1}^M \frac{\phi_p}{N_p}
\ln \phi_p + \frac{2 \pi \lB \rho_{\rm tot} }{\kappa_{\rm D}^2} 
\left( \sum_{p=1}^M
\frac{q_p}{N_p} \phi_p \right)^2 \\
& \hspace{2.0cm} + \frac{1}{\rho_{\rm tot}} \int
\frac{\dd \kk }{(2 \pi)^3} \ln\left(1 + \frac{4 \pi \lB 
\rho_{\rm tot}}{\kappa_{\rm D}^2 + \kk^2} \sum_{p=1}^M \phi_p g_p(|\kk|) \right) \; ,
\end{aligned}
\end{equation}
which is also provided as Eq.~6 in the maintext.
The free energy is expressed using volume fractions $\phi_p = N_p n_p / V
\rho_{\rm tot}$ and $\phi_{\rm w} = n_{\rm w}/V \rho_{\rm tot}$, where the
solvent volume fraction should be treated as function of the polymeric
component volume fractions, $\phi_{\rm w} = 1 - \sum_{p=1}^M \phi_p$, due to
incompressibility.

The free energy $f_{\rm RPA}$ contains a mean-field contribution from the
screened electrostatic interactions that depends on the net electric charge of
the polymer chains, $q_p \equiv \sum_{\alpha=1}^{N_p} \sigma_{p,\alpha}$.
All sequence dependence in this formulation 
originates from the charge-charge correlation functions
\begin{equation} \label{eq:g_function}
g_p(k) = \frac{1}{N_p} \sum_{\alpha=1}^{N_p} \sum_{\beta=1}^{N_p}
\sigma_{p,\alpha} \sigma_{p,\beta} \e^{-|\alpha - \beta| k^2 b^2 / 6 - k^2
a_{\rm s}^2} \; , \end{equation}
which follow from the quadratic expansion of the single polymer partition
functions $Q_p$.

Construction of co-existence curves follows from matching the exchange
chemical potentials $\bar{\mu}_i = {\partial f_{\rm RPA}}/{\partial
\phi_i}$ and osmotic pressure $\bar{\Pi} = 
\sum_i \bar{\mu}_i \phi_i - f_{\rm RPA}$ across every co-existing phase. 
Explicit expressions for $\bar{\mu}_i$
and $\bar{\Pi}$ are given by
Eqs.~\eqref{eq:rpa_chem_pot} and \eqref{eq:rpa_pressure} below.  \\
\\

\noindent{\bf The field-theory representation of multiple-chain 
polymer systems}

We now briefly demonstrate the equivalence between the system described by
the explicit-chain (particles in coordinate space) Hamiltonian $\hat{H}$ in
Eq.~\eqref{eq:particle_hamiltonian} 
and the statistical field theory defined by
the partition function in Eq.~\eqref{eq:field_partition_function}.
The procedure for converting the
particle-based degrees of freedom ($\rr_i$ and $\RR_{p,i,\alpha}$) to
fluctuating fields ($\eta(\rr)$ and $\psi(\rr)$) is documented in more detail
elsewhere\cite{Edwards1965,Fredrickson2002,Fredrikson2023_book,MiMB2023}. The
starting point is the statistical mechanical partition function written as
integrals over the positions of the molecular components, 
\begin{equation}
\label{eq:particle_partition_func}
Z = \frac{1}{n_{\rm w}! \prod_{p=1}^M n_p!} \int \lbrace \dd \rr \rbrace \lbrace
\dd \RR \rbrace \e^{-\hat{H}} 
\end{equation}
where the integration measures run over positions of all solvent- and polymer
beads in the system, $\lbrace \dd \rr \rbrace \equiv \prod_{i=1}^{n_{\rm w}}
\dd \rr_i$ and $\lbrace \dd \RR \rbrace \equiv \prod_{p=1}^{M}
\prod_{i=1}^{n_p} \prod_{\alpha=1}^{N_p} \dd \RR_{p,i,\alpha} $. Throughout
the derivation that follows, we ignore multiplicative constants in $Z$ of no
thermodynamic consequence, i.e.,~constants that provide, at most,
density-independent shifts to the free energy, chemical potentials or
pressures.

The terms in $\hat{H}$ for excluded-volume and electrostatic interactions can
be linearized in densities using Hubbard-Stratonovic transformations that
introduces the number- and charge-density conjugate fields $\eta(\rr)$ and
$\psi(\rr)$, 
\begin{equation} \label{eq:HS_transformations}
\begin{aligned}
\e^{-\int \dd \rr \hat{\rho}^2 / 2 \gamma_{\rm ex}} &\propto \int \DD \eta \, \e^{-\int
\dd \rr ( \gamma_{\rm ex} \eta^2/2 - \ii \eta \hat{\rho})}, \\ \e^{-\int \dd \rr \dd
\rr' \hat{c} \hat{c}' V_{\rm C}(\Delta r)/2 } &\propto \int \DD \psi \,
\e^{-\int \dd \rr ( \psi V_{\rm C}^{-1} \psi - \ii \psi \hat{c})}
\end{aligned}
\end{equation}
where $\hat{\rho}(\rr) = \hat{\rho}_{\rm w}(\rr) + \sum_{p=1}^M
\hat{\rho}_p(\rr)$ is the total number density at $\rr$,
$\hat{c}=\hat{c}(\rr)$ and $\hat{c}' = \hat{c}(\rr')$ denote charge densities
at $\rr$ and $\rr'$, $\Delta r = |\rr - \rr'|$, $V_{\rm C}(r)= 
\lB \e^{-\kappa_{\rm D} r}/r $ 
is the screened Coulomb potential and $V_{\rm C}^{-1} = (\kappa_{\rm D}^2 -
{\nabla}^2)/4 \pi \lB$ is the Green's function associated with the screened
Coulomb potential.

We can now substitute the definitions of the microscopic densities in
Eqs.~\eqref{eq:micro_density} and \eqref{eq:micro_charge_density} into
Eqs.~\eqref{eq:HS_transformations}. The integrals over $\lbrace \dd \rr
\rbrace$ and $\lbrace \dd \RR \rbrace$ now factorize into partition functions
for single molecules experiencing external chemical and electrostatic
potential fields $\ii \breve{\eta}$ and $\ii \breve{\psi}$, 
\begin{equation}
\begin{aligned}
Q_{\rm w}[\breve{\eta}] =& \frac{1}{V} \int \dd \rr \, 
\e^{-\ii \breve{\eta}(\rr)} , \\
Q_{p}[\breve{\eta},\breve{\psi}] =& \frac{\int \prod_{\alpha=1}^{N_p} \dd
\RR_{\alpha} }{V \mathcal{N}^{N_p-1}} \, \e^{-\frac{3}{2b^2}\sum {\Delta
\RR_{\alpha}}^2 - \sum \ii \breve{W}_{p,\alpha}} \; . \end{aligned}
\end{equation}
The single-polymer partition functions $Q_p$ contain integrals over the
positions $\RR_{\alpha}$ of the $N_p$ beads constituting the chain backbone.
In the exponent of the integrand in $Q_p$, we have denoted $\sum {\Delta
\RR_{\alpha}}^2 \equiv \sum_{\alpha=1}^{N_p-1}
(\RR_{\alpha+1}-\RR_{\alpha})^2$ and $ \sum \ii \breve{W}_{p,\alpha} =
\sum_{\alpha=1}^{N_p} \ii \breve{W}_{p,\alpha}(\RR_{\alpha})$, with
$\breve{W}_{p,\alpha}(\rr) = \breve{\eta}(\rr) + \sigma_{p,\alpha}
\breve{\psi}(\rr)$, for brevity. We have divided the single molecule partition
functions by volume factors of $V$ 
and $\mathcal{N}=(2\pi b^2/3)^{3/2}$ such that they are normalized 
to $Q_{{\rm w},p}=1$
at $\eta=\psi=0$. These volume factors in $Q_{{\rm w},p}$ are compensated by
an over-all multiplicative factor $V^{n_{\rm w}+\sum_p n_p}$ in $Z$ that,
together with the factorial factors in the denominator, corresponds to
partition function of ideal gas of (i.e.~non-interacting) polymers and solvent
molecules.

After performing the Hubbard-Stratonovich transformations and normalizing the
$Q_{{\rm w},p}$ factors, the partition function defined in
Eq.~\eqref{eq:particle_partition_func} becomes 
\begin{equation}
Z = \frac{V^{\nw+\sum_p n_p} }{\nw! \prod_p n_p!} \int \DD \eta \int \DD\psi
\, {Q_{\rm w}}^{n_{\rm w}} \left( \prod_{p=1}^M {Q_p}^{n_p} \right)  
\e^{-\int \dd \rr [ \eta^2/2\gamma_{\rm ex} + 
\psi (\kappa_{\rm D}^2-{\nabla}^2)\psi/8 \pi
\lB]}
\end{equation}
Exponentiating the $Q_{{\rm w},p}$ factors leads to our final form of the
field theory partition function in Eq.~\eqref{eq:field_partition_function}
with Hamiltonian $H[\eta,\psi]$ shown in Eq.~\eqref{eq:field_hamiltonian}. 
\\

\noindent{\bf Derivation of the free energy density in the approximate
analytical approach}

We next detail how the free energy $f_{\rm RPA}$ in
Eq.~\eqref{eq:rpa_free_energy} follows from approximate evaluations of the
functional integrals in the field theory partition function $Z$ in
Eq.~\eqref{eq:field_partition_function}. As we will ultimately consider the
$\gamma_{\rm ex} \rightarrow 0$ limit in which density-fluctuations vanish, 
we will only consider the mean-field solution for the density-conjugate field
$\eta(\rr)$ by setting $\eta(\rr)=\bar{\eta}$ where $\bar{\eta}$ solves
$\delta H / \delta \eta = 0$ for spatially homogeneous field configurations.
This gives 
\begin{equation}
\ii \bar{\eta} = \frac{1}{\gamma_{\rm ex}} 
\left( \sum_{p=1}^M \rho_p N_p + \rho_{\rm w} \right)  \; ,
\end{equation}
where $\rho_p = n_p/V$ and $\rho_{\rm w} = n_{\rm w}/V$. Similarly, the
mean-field solution for the charge-conjugate field, satisfying $\delta H/
\delta \psi = 0$ at spatially homogeneous fields, is given by
\begin{equation}
\ii \bar{\psi} = \frac{4 \pi \lB}{\kappa_{\rm D}^2} 
\sum_{p=1}^{M} q_p \rho_p \; ,
\end{equation}
where $q_p$ is the net electric charge of a $p$-type chain. The
mean-field solution $\bar{\psi}$ lacks charge-sequence dependence which only
appear at the fluctuation level of $\psi(\rr)$. We therefore write $\psi(\rr)
= \bar{\psi} + \tilde{\psi}(\rr)$, where $\tilde{\psi}(\rr)$ represents
fluctuations about $\bar{\psi}$, and expand $H$ in powers of
$\tilde{\psi}(\rr)$. In this work, we truncate this series beyond quadratic
order, corresponding to the random phase approximation (RPA), leading to
\begin{equation}
H[\eta,\psi] \approx V \bar{h} + \frac{1}{2} \int \frac{\dd \kk}{(2 \pi)^3}
G(\kk^2) \tilde{\psi}_{-\kk} \tilde{\psi}_{\kk} \; , 
\end{equation}
where $\tilde{\psi}_{\kk}$ is the Fourier transformed $\tilde{\psi}(\rr)$. The
coefficients of the quadratic term in $\tilde{\psi}$ are
\begin{equation}
G(\kk^2) = \sum_{p=1}^M g_{p}(|\kk|) N_p \rho_p + 
\frac{\kappa_{\rm D}^2 + \kk^2}{4 \pi \lB}
\; .
\end{equation}
The functions $g_{p}(k)$ (where $k\equiv |\kk|$) are provided
in Eq.~\eqref{eq:g_function} but with
$g_p(0)=0$. The property $g_p(0)=0$ does not affect the free energy and its
derivatives since the integrand of the RPA integral is zero at $\kk={\bf 0}$
because of the integration measure $\dd \kk \rightarrow 4 \pi k^2 \dd k$. The
contribution from the mean-field solutions $\bar{\eta}$ and $\bar{\psi}$ is
$H[\bar{\eta},\bar{\psi}]= V \bar{h}$ with 
\begin{equation}
\bar{h} = \frac{ \left( \sum_{p=1}^M N_p \rho_p + \rho_{\rm w} \right)^2 }{2
\gamma_{\rm ex}} 
+  \frac{2 \pi \lB}{\kappa_{\rm D}^2} \left( \sum_{p=1}^M q_p \rho_p
\right)^2 .  
\end{equation}
Evaluating the functional integrals in our approximation scheme gives
\begin{equation}
\int \DD \eta \int \DD \psi \, \e^{-H[\eta,\psi]} \approx \e^{-V (\bar{h} +
f_{\tilde{\psi}}) } \; , 
\end{equation}
where
\begin{equation}
f_{\tilde{\psi}} = \frac{1}{2} \int \frac{\dd \kk}{(2\pi)^3} \ln G(\kk^2) 
\end{equation}
comes from the Gaussian integrals over $\tilde{\psi}_{\kk}$. Note that while
the integral over $\kk$ formally diverges, this ultraviolet (UV) divergence is
independent of the polymer densities $\rho_p$ since $G(\kk^2)\rightarrow
\kk^2/4 \pi \lB$ as $|\kk| \rightarrow \infty$. The $|\kk| \gtrsim {a_{\rm
s}}^{-1}$ fluctuation modes therefore contribute only as a divergent, but
density-independent, constant to the pressure. We can remove this constant by
subtracting the fluctuation contribution at zero polymer densities,
$f_{\tilde{\psi}} \rightarrow f_{\tilde{\psi}} - 
f_{\tilde{\psi}}(\rho_1,\rho_2, \dots, \rho_M=0)$, yielding
\begin{equation}
f_{\tilde{\psi}} = \int \frac{\dd \kk}{2(2\pi)^3} \ln \left( 1 + \frac{4 \pi
\lB}{\kappa_{\rm D}^2 + \kk^2}   \sum_{p=1}^M g_{p}(|\kk|) N_p \rho_p \right) 
\end{equation}
which is finite.

The free energy density at non-zero $\gamma_{\rm ex}$ 
is $f = - V^{-1} \ln Z
\approx \sum_i \rho_i \ln \rho_i + \bar{h} + f_{\tilde{\psi}}$, where the sum
over $i$ is over all system components $i={\rm w},1,\dots,M$ and the
logarithms come from applying Sterling's approximation on the factorial
factors in Eq.~\eqref{eq:particle_partition_func}, $\ln n! \approx n \ln n$.
Here, $f(\rho_{\rm w}, \rho_1, \dots, \rho_M)$ is expressed as a function of
the number densities $\lbrace \rho_{\rm w}, \rho_1, \dots, \rho_M \rbrace$
which are independent variables at non-zero $\gamma_{\rm ex}$. 
At $\gamma_{\rm ex} \neq 0$,
co-existing phases can be found by matching the solvent and polymer chemical
potentials $\mu_i = \partial f / \partial \rho_i$, with $i={\rm w},1,\dots,M$,
and the hydrostatic pressure $\Pi = \sum_i \rho_i \mu_i - f$ across all
phases.

Given that we are interested in the $\gamma_{\rm ex} \rightarrow 0$ limit, 
a more convenient set of independent variables is $\lbrace \rho_{\rm
tot},\rho_1,\dots,\rho_M\rbrace$, which is obtained by replacing the solvent
density $\rho_{\rm w}$ as an independent variable with the 
total number density $\rho_{\rm tot}
= \rho_{\rm w} + \sum_{p=1}^M \rho_p$. By defining 
\begin{equation}
f(\rho_{\rm w}, \rho_1, \dots, \rho_M) = \bar{f}(\rho_{\rm tot}, \rho_1,
\dots, \rho_M) , 
\end{equation}
one can show that matching of the $M+1$ chemical potentials $\mu_i$ and the
hydrostatic pressure $\Pi$ is equivalent to matching the $M$ \textit{exchange}
chemical potentials $\bar{\mu}_p = \partial \bar{f} / \partial \rho_p$, the
``total'' chemical potential $\mu_{\rm tot} = \partial \bar{f}/ \partial
\rho_{\rm tot}$ and the \textit{osmotic} pressure $\bar{\Pi} = \sum_{p=1}^M
\rho_p \bar{\mu}_p - \bar{f}$. These quantities are related by
\begin{equation}
\begin{aligned}
\mu_p &= \bar{\mu}_p + N_p \mu_{\rm tot} , \quad p=1,\dots,M \; , \\
\mu_{\rm w} &= \mu_{\rm tot} \; , \\
\Pi &= \bar{\Pi} + \rho_{\rm tot} \mu_{\rm tot} \; .
\end{aligned}
\end{equation}
At small $\gamma_{\rm ex}$, $\mu_{\rm tot}$ is dominated by the 
excluded-volume interactions, viz., 
\begin{equation}
\mu_{\rm tot} \approx \frac{\rho_{\rm tot}}{\gamma_{\rm ex}} 
\quad\quad\quad \mbox{for small
$\gamma_{\rm ex}$}.  \end{equation}
Equal $\mu_{\rm tot}$ in co-existing phases then dictates that the total
density $\rho_{\rm tot}$ is the same in every phase as 
$\gamma_{\rm ex} \rightarrow 0
$. The emerging incompressibility is made manifest by replacing number
densities by volume fractions, $\phi_p = \rho_p N_p / \rho_{\rm tot}$ and
$\phi_{\rm w} = 1-\sum_{p=1}^M \phi_p$, and treating $\rho_{\rm tot}$ as
constant. Furthermore, we can neglect the 
$\rho_{\rm tot}^2/2\gamma_{\rm ex}$ term in
$\bar{h}$ for incompressible systems since it contributes equally to every
phase. We further normalize the free energy by dividing $\bar{f}$ by
$\rho_{\rm tot}$ to arrive at the dimensionless Flory-Huggins-type free energy
$f_{\rm RPA}$ given in Eq.~\eqref{eq:rpa_free_energy}.

The exchange chemical potentials and the osmotic pressure can next be
re-defined using derivatives w.r.t.~$\phi_p$ rather than $\rho_p$ (yielding
equivalent expressions up to trivial normalization factors of $N_p$). Using
$\bar{\mu}_p = \partial  f_{\rm RPA} / \partial \phi_p$ gives 
\begin{equation}
\label{eq:rpa_chem_pot}
\begin{aligned}
\bar{\mu}_p =& \frac{\ln\phi_p}{N_p} - \ln \phi_{\rm w} + \frac{4 \pi \lB
\rho_{\rm tot}}{\kappa_{\rm D}^2} 
\frac{q_p}{N_p} \sum_{q=1}^M \frac{q_q
\phi_q}{N_q}  \\ &+ \frac{\lB}{\pi} \int_0^{\infty} \dd k \frac{k^2
g_p(k)}{(\kappa_{\rm D}^2+k^2)(1+ \mathcal{A})}
\end{aligned}
\end{equation}
up to an irrelevant additive constant, where
\begin{equation}
\mathcal{A}(k) = 
 \frac{4 \pi \lB \rho_{\rm tot}}{\kappa_{\rm D}^2 + k^2}\sum_{p=1}^M
\phi_p g_p(k) \; . 
\end{equation}
The expression for the osmotic pressure, which follows from $\bar{\Pi} =
\sum_{p=1}^M \phi_p  \bar{\mu}_p - f_{\rm RPA}$, is 
\begin{equation}
\label{eq:rpa_pressure}
\bar{\Pi} = \sum_{p=1}^M \frac{\phi_p}{N_p} + \phi_{\rm w} - \ln\phi_{\rm w}
+ \frac{4 \pi \lB \rho_{\rm tot}}{\kappa_{\rm D}^2}\left( \sum_{p=1}^M
\frac{q_p \phi_p}{N_p} \right)^2 
+ \frac{\rho_{\rm tot}^{-1}}{4 \pi^2 } \int_0^{\infty} \dd k \, k^2 \left(
\frac{\mathcal{A}}{1+\mathcal{A}} - \ln(1+\mathcal{A}) \right) \;
\end{equation}
up to an irrelevant additive constant.
\\

\noindent{\bf Phenomenlogical phase-field treatment of sequence-dependent
surface tension}

The RPA free energy density $f_{\rm RPA}$ in Eq.~\eqref{eq:rpa_free_energy}
estimates the free energy density inside a bulk phase but does not account for
the free energy cost associated with interfaces separating the co-existing
phases. While such effects inherently vanish in the infinite-volume limit,
they can be crucial in systems of finite size. To account for surface effects,
we construct a free energy functional $F[\lbrace \rho_{p}(\rr) \rbrace]$
that extends the RPA free energy with a term that punishes density gradients
(Eq.~1 of the maintext):
\begin{equation}
F[\lbrace \rho_{p}(\rr) \rbrace] = \int \dd \rr \left[ \sum_{p,q}
\frac{\xi_{p,q}}{2} \bm{\nabla}\rho_p(\rr) \cdot \bm{\nabla}\rho_q(\rr) +
f_{\rm RPA}(\lbrace \rho_p(\rr) \rbrace) \right] 
\end{equation}
which reduces to $F \rightarrow V f_{\rm RPA}$ for spatially constant density
configurations $\rho_p(\rr) = \bar{\rho}_p \equiv  n_p N_p / V$,
$p=1,\dots,M$. In principle, the matrix $\xi_{p,q}$ (as well as information
about higher-order gradient terms) could be derived from the partition
function in Eq.~\eqref{eq:particle_partition_func} but that is beyond the 
scope of the current work. Instead, here we take a more modest approach in
which
$\xi_{p,q}$ are treated as free parameters to be adjusted to reproduce trends
observed in MD simulations.

We determine the equilibrium spatial density distributions by 
minimizing the free energy (while neglecting 
fluctuations to higher-than-lowest free energies), i.e.,~by solving 
(Eq.~7 of the maintext)
\begin{equation}
\frac{\delta F}{\delta \rho_p(\rr)} = -\sum_q \xi_{p,q} \bm{\nabla}^2
\rho_q(\rr) + \mu_p(\lbrace \rho_q(\rr) \rbrace) = 0 \;
\end{equation}
subject to the number conservation constraints $\int \dd \rr \rho_p(\rr) / V =
\bar{\rho}_p$. These constraints are automatically satisfied through the
change of variables from $\rho_p(\rr)$ to $\varphi_p(\rr)$, where
\begin{equation}
\rho_p(\rr;\varphi_p(\rr)) = \bar{\rho}_p \frac{\e^{\varphi_p(\rr)}}{\int \dd
\rr' \e^{\varphi_p(\rr')}  / V} \; .
\end{equation}
Note that the variable $\varphi_p(\rr)$ is different from, and therefore
should not be confused with the volume fraction $\phi_p$.
Now, minimizing the free energy instead with respect to~$\varphi_p(\rr)$ 
amounts to solving
\begin{equation}
\begin{aligned}
0 = \frac{\delta F}{\delta \varphi_p(\rr)} = \int \dd \rr' \frac{\delta
F}{\delta \rho_p(\rr') } \frac{\delta \rho_p(\rr')}{\delta \varphi_p(\rr)} =
\rho_p(\rr) \left[ \frac{\delta F}{\delta \rho_p(\rr)} - \frac{1}{\bar{\rho}_p
V} \int \dd \rr' \frac{\delta F}{\delta \rho_p(\rr') } \right].  \end{aligned}
\end{equation}
We numerically solve the above equation by introducing a time variable $t$ and
evolve $\varphi_p(\rr,t)$ according to 
\begin{equation}
\frac{\partial \varphi_p(\rr,t)}{\partial t} = - \frac{\delta F}{\delta
\rho_p(\rr)} +\frac{1}{\bar{\rho}_p V} \int \dd \rr' \frac{\delta F}{\delta
\rho_p(\rr') } \;
\end{equation}
until a steady-state configuration has been reached.
This formulation bears 
similarity with the Cahn-Hillard approach \cite{Cahn1958} 
in the sense that it produces an evolution towards a density distribution
that minimizes the free energy.
However, it should be noted that here the variable $t$ does not correspond
to real time and the above equation produces nonlocal changes in
$\rho(\rr)$.
\\

\vfill\eject

\noindent {\large\it Surface tension in one-component systems}

To facilitate comparison with MD results (e.g. Fig.~S6),
we now focus on a system with a single polymeric component (thus
$p$ subscripts can be dropped) and restrict our
analysis to ``slab geometry'' density configurations,
i.e.,~$\rho(\rr)=\rho(z)$, that vary only in the $z$-direction. These
restrictions allow us to compute surface tension $\gamma$, i.e.,~the free
energy excess cost per unit area of the interface in the absence of surface
curvature.

The free energy per area element perpendicular to the $z$-direction is
is now given by
\begin{equation}
\frac{F[\varphi(z)]}{L_x L_y} = \int_0^{L_z} \dd z \left[ \frac{\xi}{2} \left(
\frac{\dd \rho(z) }{\dd z} \right)^2 + f_{\rm RPA}(\rho(z)) \right],
\end{equation}
where $\rho(z) =\bar{\rho} \, \e^{\varphi(z)} \,  L_z / \int_0^{L_z} \dd z \,
\e^{\varphi(z)}$. By using 
the relation between $\rho$ and $\varphi$ 
as well as the identity
$\delta F / \delta \varphi(z)=\int \dd z' [\delta F/\delta\rho(z')] 
[\delta\rho(z')/\delta \varphi(z)]$,
the minimization condition $\delta F / \delta \varphi(z)=0$
can be rewritten as 
\begin{equation}
-\xi \frac{\dd^2 \rho(z) }{\dd z^2} + \mu(\rho(z)) = X
\end{equation}
where $X = (\int\dd z'\rho(z') [{\delta F}/{\delta \rho(z')}])/V\bar{\rho}$ 
is a constant independent of $z$. Since the chemical
potential $\mu(\rho) = \partial f_{\rm RPA} / \partial \rho$,
one can multiply the above equation with $\dd \rho(z) / \dd z$ such that both
sides of the equation become total derivatives. Integrating the resulting
expressions gives the following general condition for the equilibrium density
configuration $\rho(z)$, 
\begin{equation}
\label{eq:1d_equilibrium_profile_condition}
-\frac{\xi}{2} \left( \frac{\dd \rho(z)}{\dd z} \right)^2 + 
f_{\rm RPA}(\rho(z)) = X \rho(z) + Y , \end{equation}
for some constant $Y$. When the system phase separates 
into a dilute phase and a condensed phase, the density profile has 
spatially homogeneous
regions with $\rho(z) = \rho_{\rm d}$ (dilute) and $\rho(z) = \rho_{\rm c}$
(condensed) in which the gradient contribution to the free energy vanishes. We
can use these regions to deduce the values of $X$ and $Y$:
\begin{equation}
\begin{aligned}
f_{\rm RPA}(\rho_{\rm d} ) &= X \rho_{\rm d} + Y \\
f_{\rm RPA}(\rho_{\rm c} ) &= X \rho_{\rm c} + Y
\end{aligned}
\end{equation}
which can be solved for $X$ and $Y$. However, we can simplify the calculation
by shifting the RPA free energy by a linear relation in $\rho$
(which has no consequence on phase separation) such that 
$f_{\rm RPA}(\rho_{\rm d}) =
f_{\rm RPA}(\rho_{\rm c}) = 0$, which results in $X=Y=0$.

We are now in a position to calculate the surface tension $\gamma$ between
two co-existing phases. Consider a density profile $\rho(z)$ interpolating
between a dilute phase at $z=a$, $\rho(a)=\rho_{\rm d}$ and a condensed phase
at $z=b$, $\rho(b) = \rho_{\rm c}$. Given the relation in
Eq.~\eqref{eq:1d_equilibrium_profile_condition} combined with $X=Y=0$, the
surface tension, i.e. free energy per area, is 
\begin{equation}
\label{eq:xi_gamma_relation}
\gamma = \xi \int_a^b \dd z \, \left( \frac{\dd \rho(z)}{\dd z} \right)^2 = 2
\int_a^b \dd z \, f_{\rm RPA}(\rho(z)) \; . 
\end{equation}
The first equality gives the surface tension $\gamma$ purely in terms of the
density profile $\rho(z)$ without reference to $f_{\rm RPA}$. Alternatively,
the second equality can be used to evaluate $\gamma$ purely in terms of
$f_{\rm RPA}$ and $\rho_{\rm d,c}$ without reference to the specific shape of
$\rho(z)$ as follows. First, note that
Eq.~\eqref{eq:1d_equilibrium_profile_condition} gives $\dd \rho(z) / \dd z =
\sqrt{f_{\rm RPA}(\rho(z)) / \xi}$. Since $\rho(z)$ is a monotonically
increasing function from $\rho(a)=\rho_{\rm d}$ to $\rho(b)=\rho_{\rm c}$, we
can invert the function $\rho(z)$ to $z(\rho)$ and convert the integral over
$z$ to an integral over $\rho$ as 
\begin{equation}
\gamma = 2 \int_{\rho_{\rm d}}^{\rho_{\rm c}} \dd \rho \, \frac{\dd z(\rho)
}{\dd \rho} f_{\rm RPA}(\rho) = 2 \sqrt{\xi} \int_{\rho_{\rm d}}^{\rho_{\rm
c}} \dd \rho \, \sqrt{f_{\rm RPA}(\rho)} \;
\end{equation}
provided $f_{\rm RPA}$ has been shifted such that
$f_{\rm RPA}(\rho_{\rm d}) = f_{\rm RPA}(\rho_{\rm c}) = 0$.
\\

\noindent
{\large \it Fitting the interface coefficients}

We estimate the sequence dependence of the interface coefficients $\xi_{ij}$
by first fitting an approximate relationship between $\xi_{ii}$ and existing
MD simulation data on the surface tension in one-component systems through
Eq.~\eqref{eq:xi_gamma_relation}. This relation implies that, given values for
the one-component surface tension $\gamma_i$ for a set of sequences $p$, the
relative values of $\xi_{p,p}$ can be estimated as 
\begin{equation}
    \xi_{p,p} \propto \left( \frac{\gamma_p}{\int_{\rho_{\rm d}}^{\rho_{\rm c}}
\dd \rho \, \sqrt{f_{\rm RPA}(\rho)}} \right)^2 \, .  
\end{equation}
In Devajaran \textit{et al.}~\cite{MittalNatComm2024}, 
$\gamma_p$ was computed for
fourteen fully charged but overall electrically neutral 
50-mer E/K sv-type sequences under phase-separating
conditions. The inferred values of $\xi_{p,p}/\xi_{\rm sv30}$ against the
charge-sequence decoration (SCD) parameter are shown in Fig.~S7.
The best fit (dashed line in Fig.~S7b) is given by
\begin{equation}
    \xi_{p,p}/\xi_{\rm sv30} = 0.368 - 0.020 \cdot {\rm SCD} \, .
\end{equation}
We make contact with our own MD simulations by setting the overall scale of
$\xi_{p,p}$ such that the 1D one-component phase-field model reproduces the
interface width of the sv28 MD density profile. The interface width $d$, in
turn, in the respective models is determined by fitting the density profile of
the interface to a hyherbolic tangent function, 
\begin{equation}
    \rho(z) = \frac{1}{2}\left( \rho_{\rm c} + \rho_{\rm d} \right) +
\frac{1}{2} \left( \rho_{\rm c} - \rho_{\rm d} \right) \tanh\left( \frac{z -
z_0}{d} \right) \, .  
\end{equation}
The resulting MD sv28 interface width $d=1.42b$ yielded an overall scale
factor for $\xi_{p,p}$ as 
\begin{equation}
    \xi_{p,p} = \left(0.368 - 0.020 \cdot {\rm SCD} \right) \cdot 246 b^8 \, ,
\end{equation}
which is also presented as Eq.~2 in the maintext.
As discussed in the maintext,
the cross-sequence interface coefficients are set to
(Eq.~3 of the maintext)
\begin{equation}
\xi_{p,q} = \sqrt{\xi_{p,p} \xi_{q,q}}
\frac{\min(|\mathrm{SCD}_p|,|\mathrm{SCD}_q|)}{\max(|\mathrm{SCD}_p|,|\mathrm{SCD}_q|)}
\, , \end{equation}
since this produced reasonable interfaces in the two-component runs while
reducing to $\xi_{p,q} = \xi_{p,p} = \xi_{q,q}$ when the 
two species are the same sequence.
\\


$\null$\\
$\null$\\

\noindent{\large\bf Gibbs Ensemble-like Monte Carlo (MC) Model: 
Constrained Moves and Example of Thermal Fluctuation in a Simple System}\\

\noindent
{\bf MC moves for two-component systems}

The MC moves employed to sample the Gibbs ensemble-like system governed
by the partition function $Z[\{\bar{\phi}_p\};V_{\rm v},N_{\rm v}]$ in
maintext Eq.4 for one-component (a single $\phi$ variable) systems are 
given in {\it Materials and Methods} in the maintext.
As described, the MC moves are constrained to satisfy the overall input
volume fractions $\{\bar{\phi}_p\}$ at every step of the MC simulation and
are designed such that the sampled distribution $P(\{\phi_p^{(k)}\})\propto
\exp[-V_{\rm v}\sum_{k=1}^{N_{\rm v}} f(\{\phi_p^{(k)}\})]$ at equilibrium.
The corresponding MC moves used for two-component systems are as follows:
\begin{itemize}
\item
Let $(\phi_1,\phi_2)$ be volume fractions of the two components. 
Similar to the one component case, both $\phi_1$ and $\phi_2$ can take
200 possible values in the range $[0.005, 0.99]$.
\item Each voxel is initialized with an input $(\bar{\phi}_1, \bar{\phi}_2)$
with $\bar{\phi}_1$ and $\bar{\phi}_2$ each equals to one of the discretized
values.
\item
Two voxels $k,k^\prime$ with existing
$(\phi_1,\phi_2)=$ 
$(\phi^{(k)}_{1,\mathrm{old}},\phi^{(k)}_{2,\mathrm{old}})$, 
$(\phi^{(k^\prime)}_{1,\mathrm{old}},\phi^{(k^\prime)}_{2,\mathrm{old}})$ 
are randomly selected.
\item
A new $(\phi_{1,\mathrm{new}},\phi_{2,\mathrm{new}})$ 
is randomly selected among the discretized $(\phi_1,\phi_2)$ values
to replace
$(\phi^{(k)}_{1,\mathrm{old}},\phi^{(k)}_{2,\mathrm{old}})$.
\item
$(\phi^{(k^\prime)}_{1,\mathrm{old}},\phi^{(k^\prime)}_{2,\mathrm{old}})$
is then replaced by
$([\phi^{(k^\prime)}_{1,\mathrm{old}} - (\phi^{(k)}_{1,\mathrm{new}} -
\phi^{(k)}_{1,\mathrm{old}})], [\phi^{(k^\prime)}_{2,\mathrm{old}} - 
(\phi^{(k)}_{2,\mathrm{new}} - \phi^{(k)}_{2,\mathrm{old}})])$
such that $(\bar{\phi}_1, \bar{\phi}_2)$ is not changed.
\item
This attempted move is rejected if any component in 
$(\phi^{(k)}_{1,\mathrm{new}},\phi^{(k)}_{2,\mathrm{new}})$.
or
$(\phi^{(k^\prime)}_{1,\mathrm{new}},\phi^{(k^\prime)}_{2,\mathrm{new}})$.
is $<0$ or $>1$. The move is also rejected if
$\phi^{(k)}_{1,\mathrm{new}}+\phi^{(k)}_{2,\mathrm{new}}>1$ or
$\phi^{(k^\prime)}_{1,\mathrm{new}}+\phi^{(k^\prime)}_{2,\mathrm{new}}>1$.
\item Let $\Delta f=$
$f(\phi^{(k)}_{1,\mathrm{new}},\phi^{(k)}_{2,\mathrm{new}})+
f(\phi^{(k^\prime)}_{1,\mathrm{new}},\phi^{(k^\prime)}_{2,\mathrm{new}})-
f(\phi^{(k)}_{1,\mathrm{old}},\phi^{(k)}_{2,\mathrm{old}})-
f(\phi^{(k^\prime)}_{1,\mathrm{old}},\phi^{(k^\prime)}_{2,\mathrm{old}})$.
The attempted move is accepted if a random number $\in [0,1]$ is
$\leq e^{-V_v \Delta f}$.
\end{itemize}
For maintext Fig.~7e--g,
$f$ is RPA free energy in this work; for Supporting Fig.~S13,
$f$ is the two-component FH free energy for 50-bead chains with
Flory $\chi_{11}=\chi_{22}=0.66$ and $\chi_{12}=0.33$ that
allows for ternary LLPS (Fig.~7 of ref.~\citen{njp2017}). 
\\

\noindent{\bf Sampling}

For the one-component system studied in the maintext,
equilibration of the MC simulation is monitored by plotting the number 
of attempted moves versus the free energy density.
Once equilibration is achieved, $P(\phi)$ histograms are sampled
from the instantaneous $\phi$ values of the voxels 
at a regular interval of $10^5$ attempted moves, binning the counts
into the $200$ bins in the $[0,1]$ intervals, i.e., the number of bins 
is equal to the number of discretized $\phi$ values.
The results shown in maintext Fig.~7b--d are the normalized histogram
averaged over $2,000$ samples.

For two-component systems at higher $V$ values, to overcome difficulties
in achieving equilibration, we first run a long simulation for 
$V=2.5 \times 10^3 b^3$ for each sequence pair such that the systems are 
well equilibrated. To accelerate simulations of
reliable samples for $V = 5\times 10^3 b^3$, $2.5\times 10^4 b^3$
and $5\times 10^4 b^3$, we start the simulations for the latter
$V$ values  by initializing the voxels with
$(\phi_1,\phi_2)$ values at the last attempted move of the  
 $V=2.5\times 10^3 b^3$ simulations. Once equilibration is
achieved, we sampled $P(\phi_1, \phi_2)$ histograms from the 
instantaneous $(\phi_1, \phi_2)$ values of the cells into 
$200 \times 200$ bins at a regular interval of $10^5$ attempted moves. 
The results shown in maintext Fig.~7e--g are the normalized histogram 
averaged over $6,000$ samples. 
\\

\noindent{\bf Application of the model to a simple homogeneous system}

To gain better understanding of the partition function formulation in 
maintext Eq.~4 which is applicable to situations with an arbitrary 
number of coexisting phases (maintext Eq.~5), it is instructive---and
also a useful test---to apply the formulation to a one-component
homogeneous system governed by a general free energy density 
$f(\phi)$ [in units of $k_{\rm B}T$] as a function of the volume fraction 
$\phi$ of the single component.

Consider the following continuum form of Eq.~4 for $f(\phi)$:
\begin{equation}
\label{eq:Zphi}
\begin{aligned}
Z[\phi;V_{\rm v},N_{\rm v}] = & \int_0^1 \dd\phi_1 
\int_0^1 \dd\phi_2 \; \dots
\int_0^1 \dd\phi_{N_{\rm v}}
\exp \left ( -V_{\rm v}\left [ f(\phi_1)+f(\phi_2)+\dots +f(\phi_{N_{\rm v}})
\right ]\right) \\
& \hspace{5cm}
\times \delta \left ( 
\bar{\phi} -[\phi_1+\phi_2+\dots +\phi_{N_{\rm v}}]/{N_{\rm v}} \right )
\; .
\end{aligned}
\end{equation}
To evaluate $Z[\phi;V_{\rm v},N_{\rm v}]$ approximately, we
expand $f(\phi)$s around $\bar{\phi}$, viz.,
$f(\phi_k)=f(\bar{\phi})+(\phi_k-\bar{\phi})f^\prime(\bar{\phi})
+(\phi_k-\bar{\phi})^2f^{\prime\prime}(\bar{\phi})/2! +\dots$ for $k=1,2,\dots,
N_{\rm v}$,
where the prime superscripts represent derivatives with respect to $\phi_k$.
Because the zeroth order term is independent of 
the integration variables $\phi_1,\phi_2,\dots,\phi_{N_{\rm v}}$ and
the sum of first order terms $\propto (\phi_1-\bar{\phi}) +
(\phi_2-\bar{\phi}) + \dots +(\phi_{N_{\rm v}}-\bar{\phi})$
vanishes because of the $\delta$-function in Eq.~\ref{eq:Zphi},
\begin{equation}
\begin{aligned}
Z[\phi;V_{\rm v},N_{\rm v}] \approx & \;\; \e^{-Vf(\bar{\phi})}
\int_0^1 \dd\phi_1 \;
\e^{-V_{\rm v}(\phi_1-\bar{\phi})^2f^{\prime\prime}(\bar{\phi_1})/2}
\int_0^1 \dd\phi_2 \; 
\e^{-V_{\rm v}(\phi_2-\bar{\phi})^2f^{\prime\prime}(\bar{\phi_2})/2}
\dots \\
& \dots\times
\int_0^1 \dd\phi_{N_{\rm v}} \;
\e^{-V_{\rm v}(\phi_{N_{\rm v}}-\bar{\phi})^2
f^{\prime\prime}(\bar{\phi}_{N_{\rm v}})/2}
\;
\delta \left ( 
\bar{\phi} -[\phi_1+\phi_2+\dots +\phi_{N_{\rm v}}]/{N_{\rm v}} \right )
\; ,
\end{aligned}
\end{equation}
where $V=N_{\rm v}V_{\rm v}$ as in the maintext.
Using the Fourier representation
\begin{equation}
\delta \left ( \bar{\phi} -[\phi_1+\phi_2+\dots +\phi_{N_{\rm v}}]/{N_{\rm v}}
\right )
=\frac {N_{\rm v}}{2\pi}\int_{-\infty}^{\infty} \dd\omega \;
\e^{\ii\omega [(\bar{\phi}-\phi_1) + \ii\omega (\bar{\phi}-\phi_2) +\dots +
\ii\omega (\bar{\phi}-\phi_{N_{\rm v}})]}
\end{equation}
and changing the order of $\int\dd\omega$ and the $\prod_k\int\dd\phi_k$
integrations lead to
\begin{equation}
\label{eq:Zphiexp}
Z[\phi;V_{\rm v},N_{\rm v}] \approx 
\frac {N_{\rm v}\e^{-Vf(\bar{\phi})}}{2\pi}\int_{-\infty}^{\infty} \dd\omega
\prod_{k=1}^{N_{\rm v}}\int_0^1 \dd\phi_k \;
\e^{-V_{\rm v}(\phi_k-\bar{\phi})^2f^{\prime\prime}(\bar{\phi})/2
+\ii\omega [(\bar{\phi}-\phi_k)} \\
\; .
\end{equation}
If the Gaussian integrands in the above $\int_0^1 \dd\phi_k$ integrations
are sufficiently narrow, i.e., 
$V_{\rm v}f^{\prime\prime}(\bar{\phi})\gg 1$, the $\int_0^1 \dd\phi_k$
integrals may be approximated by $\int_{\infty}^\infty \dd\phi_k$ integrals.
Since
\begin{equation}
\int_{-\infty}^\infty \dd\phi_k \;
\e^{-V_{\rm v}(\phi_k-\bar{\phi})^2f^{\prime\prime}(\bar{\phi})/2
+\ii\omega [(\bar{\phi}-\phi_k)} =
\sqrt{\frac {2\pi}{f^{\prime\prime}(\bar{\phi})}}
\e^{-\omega^2/2V_{\rm v}f^{\prime\prime}(\bar{\phi})}
\end{equation}
is independent of $k$, Eq.~\ref{eq:Zphiexp} becomes
\begin{equation}
\begin{aligned}
Z[\phi;V_{\rm v},N_{\rm v}] & \approx 
\frac {N_{\rm v}\e^{-Vf(\bar{\phi})}}{2\pi}
\left ( \frac {2\pi}{f^{\prime\prime}(\bar{\phi})} \right )^{N_{\rm v}/2}
\int_{-\infty}^{\infty} \dd\omega
\e^{-\omega^2 N_{\rm v}/2V_{\rm v}f^{\prime\prime}(\bar{\phi})} \\
&
=\sqrt{Vf^{\prime\prime}(\bar{\phi})} \;
\e^{-Vf(\bar{\phi})}
\left ( \frac {2\pi}{f^{\prime\prime}(\bar{\phi})} \right )^{N_{\rm v}/2}
\; .
\end{aligned}
\end{equation}
Now consider
the variance of the volume fractions of the voxels 
$\langle (\phi-\bar{\phi})^2 \rangle$ defined as
$N_{\rm v}^{-1}\sum_{k=1}^{N_{\rm v}} \langle (\phi_k-\bar{\phi})^2 \rangle$,
which is equal to $\langle (\phi_1-\bar{\phi})^2 \rangle$
because of the symmetry among $\phi_k$s. The latter
is evaluated by using the relation
\begin{equation}
\begin{aligned}
\int_0^1 \dd\phi_1 \; &
(\phi_1-\bar{\phi})^2
\e^{-V_{\rm v}(\phi_1-\bar{\phi})^2f^{\prime\prime}(\bar{\phi_1})/2
+\ii\omega [(\bar{\phi}-\phi_1)} \approx
\int_{-\infty}^\infty \dd\phi_1 \; 
(\phi_1-\bar{\phi})^2
\e^{-V_{\rm v}(\phi_1-\bar{\phi})^2f^{\prime\prime}(\bar{\phi_1})/2
+\ii\omega [(\bar{\phi}-\phi_1)}\\
= & 
-\frac {\partial^2}{\partial\omega^2}
\sqrt{\frac {2\pi}{f^{\prime\prime}(\bar{\phi})}}
\e^{-\omega^2/2V_{\rm v}f^{\prime\prime}(\bar{\phi})}
= 
\sqrt{\frac {2\pi}{f^{\prime\prime}(\bar{\phi})}}
\left ( \frac {1}{V_{\rm v}f^{\prime\prime}(\bar{\phi})} - 
\frac {\omega^2}{V^2_{\rm v}f^{\prime\prime}(\bar{\phi})^2} \right )
\e^{-\omega^2/2V_{\rm v}f^{\prime\prime}(\bar{\phi})}
\end{aligned}
\end{equation}
in the final overall $\int \dd\omega$ integration to yield
\begin{equation}
\begin{aligned}
\langle (\phi-\bar{\phi})^2 \rangle =
\langle (\phi_1-\bar{\phi})^2 \rangle & \approx 
\frac {1}{Z[\phi;V_{\rm v},N_{\rm v}]}
\frac {N_{\rm v}\e^{-Vf(\bar{\phi})}}{2\pi}
\left ( \frac {2\pi}{f^{\prime\prime}(\bar{\phi})} \right )^{N_{\rm v}/2}
\\
& \quad \times
\int_{-\infty}^{\infty} \dd\omega
\left ( \frac {1}{V_{\rm v}f^{\prime\prime}(\bar{\phi})} - 
\frac {\omega^2}{V^2_{\rm v}f^{\prime\prime}(\bar{\phi})^2} \right )
\e^{-\omega^2 N_{\rm v}/2V_{\rm v}f^{\prime\prime}(\bar{\phi})}
\\
& = \frac {1}{V_{\rm v}f^{\prime\prime}(\bar{\phi})}
\left ( 1 - \frac {1}{N_{\rm v}} \right )
\xrightarrow[N_{\rm v}\to\infty]{} 
\frac {1}{V_{\rm v}f^{\prime\prime}(\bar{\phi})}
\; .
\end{aligned}
\end{equation}
The last $N_{\rm v}\rightarrow\infty$ result is consistent with
general expressions for thermal fluctuation in volume $V_{\rm v}$. 
For instance, in the $\mu VT$ ensemble (here $\mu$ stands for chemical 
potential), density fluctuation in volume $V$ is given by
$\langle (\rho-\bar{\rho})^2\rangle = 
k_{\rm B}T[V(\partial\mu/\partial\rho)_T]^{-1}$ (ref.~\cite{Mishin}).
This formula is equivalent to our expression 
$(V_{\rm v}f^{\prime\prime})^{-1}$ above when $V_{\rm v}$ is the volume of 
interest because $\rho\propto\phi$ in our models and thus the formula
applies when all the $\rho$s are replaced by $\phi$s and
since free energy density is $k_{\rm B}T f(\phi)$ in our model,
$\mu=k_{\rm B}T(\partial f/\partial\rho)_T$.
As mentioned in the maintext, for a homogeneous gas of beads with 
excluded volume but non-interacting otherwise (FH theory with $\chi=0$),
$f(\phi)=\phi\ln\phi+(1-\phi)\ln(1-\phi)$, thus
$f^{\prime\prime}(\bar{\phi})=[\bar{\phi}(1-\bar{\phi})]^{-1}$
and hence $\langle (\phi-\bar{\phi})^2 \rangle \approx 
\bar{\phi}(1-\bar{\phi})/V_{\rm v}$, which we have
verified numerically by MC simulation of our model.
\\

$\null$\\

%
\DeclareRobustCommand{\citen}[1]{%
  \begingroup
    \romannumeral-`\x 
    \setcitestyle{numbers}%
    \cite{#1}%
  \endgroup
}
%

\setcounter{table}{0}
\renewcommand{\tablename}{{\bf Table}}
\renewcommand{\thetable}{{\bf S}{\bf \arabic{table}}}

 \begin{table*}
\centerline{\Large\bf Supporting Table}
\vspace{0.1cm}
 \caption{All 50-bead ($N_p=50$) charged sequences studied in this work.
Following previous notation \cite{rohit2013,kings2015,Paletal_JPCL2024},
positively and negatively charged beads 
are symbolized by ``K'' (lysine, charge $+1$)
and ``E'' (aspartic acid, charge $-1$) respectively.
Sequence charge pattern is quantified by sequence charge decoration 
($\mbox{SCD}$) (ref.~\citen{kings2015})
and ``blockiness'' parameter $\kappa$ (ref.~\citen{rohit2013});
$\sigma_p$ is the net charge of the sequence. 
The sv sequences are from Das and Pappu \cite{rohit2013}; 
the ``constant-$\kappa$'' (c$\kappa$) and ``constant-SCD'' (cSCD) polyampholyte
sequences are from Pal et al. \cite{Paletal_JPCL2024}.
Introduced here are the ``cp'' (positively charged) `and ``cn'' (negatively charged)
sequences with net charges $q_p$s (right-most column).
Net charge per residue (NCPR) of a sequence is given by NCPR = $q_p/N_p$.
\vspace{0.3cm}
\label{tab:sequences} }

 \begin{tabular}{lllll}
 Name & Sequence & $\mbox{SCD}$ & $\kappa$ & $q_p$  \\  \hline \hline
 sv1   & \texttt{EKEKEKEKEKEKEKEKEKEKEKEKEKEKEKEKEKEKEKEKEKEKEKEKEK} & $-0.413$ & 0.000871 & 0 \\
 sv2   & \texttt{EEEKKKEEEKKKEEEKKKEEEKKKEEEKKKEEEKKKEEEKKKEEEKKKEK} & $-1.01$ & 0.00254 & 0 \\
 sv6   & \texttt{EEEKKEKKEEKEEKKEKKEKEEEKKKEKEEKKEEEKKKEKEEEEKKKKEK} & $-0.981$ & 0.0271 & 0 \\
 sv9   & \texttt{EEKKEEEKEKEKEEEEEKKEKKEKKEKKKEEKEKEKKKEKKKKEKEEEKE} & $-2.08$ & 0.0622 & 0 \\
 sv10 & \texttt{EKKKKKKEEKKKEEEEEKKKEEEKKKEKKEEKEKEEKEKKEKKEEKEEEE} & $-2.10$ & 0.0832 & 0 \\
 sv14 & \texttt{EKKEKEEKEEEEKKKKKEEKEKKEKKKKEKKKKKEEEEEEKEEKEKEKEE} & $-2.28$ & 0.131 & 0 \\
 sv18 & \texttt{KEEKKEEEEEEEKEEKKKKKEKKKEKKEEEKKKEEKKKEEEEEEKKKKEK} & $-2.04$ & 0.167 & 0 \\
 sv21 & \texttt{EEEEEEEEEKEKKKKKEKEEKKKKKKEKKEKKKKEKKEEEEEEKEEEKKK} & $-4.08$ &  0.273 & 0 \\
 sv22 & \texttt{KEEEEKEEKEEKKKKEKEEKEKKKKKKKKKKKKEKKEEEEEEEEKEKEEE} & $-4.52$ & 0.322 & 0 \\
 sv26 & \texttt{KEEEEEEEKEEKEEEEEEEEEKEEEEKEEKKKKKKKKKKKKKKKKKKKKE} & $-16.21$ & 0.610 & 0 \\
 sv29 & \texttt{KEEEEKEEEEEEEEEEEEEEEEEEEEEKKKKKKKKKKKKKKKKKKKKKKK} & $-22.57$ & 0.876 & 0 \\
 sv30 & \texttt{EEEEEEEEEEEEEEEEEEEEEEEEEKKKKKKKKKKKKKKKKKKKKKKKKK} & $-27.84$ & 1.0 & 0  \vspace{0.2cm} \\
 c$\kappa$1    & \texttt{KKKEEEEEEEEKKKKKKKKKKEEEEEEEEEEKKKKKKKKKEEKEEEKEKE} & $-2.55$ & 0.500 & 0 \\
 c$\kappa$2    & \texttt{KKKEEEEEEEKKEEEEEKKKEKKKKKKKKEEEEEEEEEEEEKKKKKKKKK} & $-4.04$ & 0.496 & 0  \\
 c$\kappa$3    & \texttt{KKKKKKKKKKKKKKKEKKKKKEKKEEEEKEEEEEKEEEEEKEEEEEEEEE} & $-19.56$ & 0.499 & 0 \\
 c$\kappa$4 & \texttt{KKKKKKKKKKEKKKKKEKKKKEKKKKEEKEEEEEKEEEEEEEEEEEEEEE} & $-19.75$ & 0.507 & 0 \vspace{0.2cm} \\
 cSCD1 & \texttt{KKKKEKKKKEEKKKKKKKEEKKKKKEEEEKKEEEEEEEKEEKEEEKEEEE} & $-9.94$ & 0.209 & 0 \\
 cSCD2 & \texttt{KEKKKKKKKKEEKKKKKKEEKKKKKEEEEEKKEEEEEEEEKKKEEEEEEE} & $-9.95$ & 0.340 & 0 \\
 cSCD3 & \texttt{EEEEEEEKKKKKKKKKKKKKKKEEEKKKKKKKKKKEEEEEEEEEEEEEEE} & $-10.03$ & 0.726 & 0\\
 cSCD4 & \texttt{EEEEEKKKKKKKKKKKKKKKKKKKKEEEEEEEEEEEEEEEEEEEEKKKKK} & $-9.75$ & 0.798 & 0   \vspace{0.2cm}  \\
 cp1   & \texttt{KKKKKKKKKKKKKKKKKKKKKKKKKKKEEEEEEEEEEEEEEEEEEEEEEE} & $-27.04$ & 0.987 & +4 \\
 cp2   & \texttt{KEKKEKKEEKEKEKEKKKEEKEEKEKKEEKKKEEKKKEEKKEEKEEKKEK} & 0.00308 & 0.000934 & +4 \\
 cp3   & \texttt{EEEKEEEEEKEEEKKKKKEKKKKKKKKEEKKKKKKKKKKKKKKKEEKEEE} & $-3.91$ & 0.378 & +12 \\
 cp4   & \texttt{KEKKEKKKEKKKKEEEKKEKKKEEKEEKKKEKEEKKKEKEKKEEKKKKEK} & 5.18 & 0.00289 & +12 \vspace{0.2cm} \\
 cn1   & \texttt{KKKKKKKKKKKKKKKKKKKKKKEKEEEEEEEEEEEEEEEEEEEEEEEEEE} & $-26.72$ & 0.947 & $-4$ \\
 cn2   & \texttt{EKEEEKEEEKEKKKKEKEEEKEKKEKEEEKEKKEEKKEEKEKEKKKKEEE} & $-0.325$ & 0.00392 & $-4$ \\
 cn3   & \texttt{KKKKKKKKKEKEEEKKKKEEEKKEEEEEEEEEEEEEEEEEEEEEKKEEKE} & $-7.30$ & 0.467 & $-12$ \\
 cn4   & \texttt{EEEEEKKKEEEKKKEEEEKEKKEKEEKKEKKEEKEEEKKEEKEEKEEEEE} & 5.20 & 0.00457 & $-12$
 \end{tabular}
 \end{table*}

\vfill\eject
\clearpage

\setcounter{figure}{0}
\renewcommand{\figurename}{{\bf Fig.}}
\renewcommand{\thefigure}{{\bf S}{\bf \arabic{figure}}}

\begin{figure}[t]
\centerline{\Large\bf Supporting Figures}
\vspace{0.4cm}
   \centering
   \includegraphics[width=0.90\columnwidth]{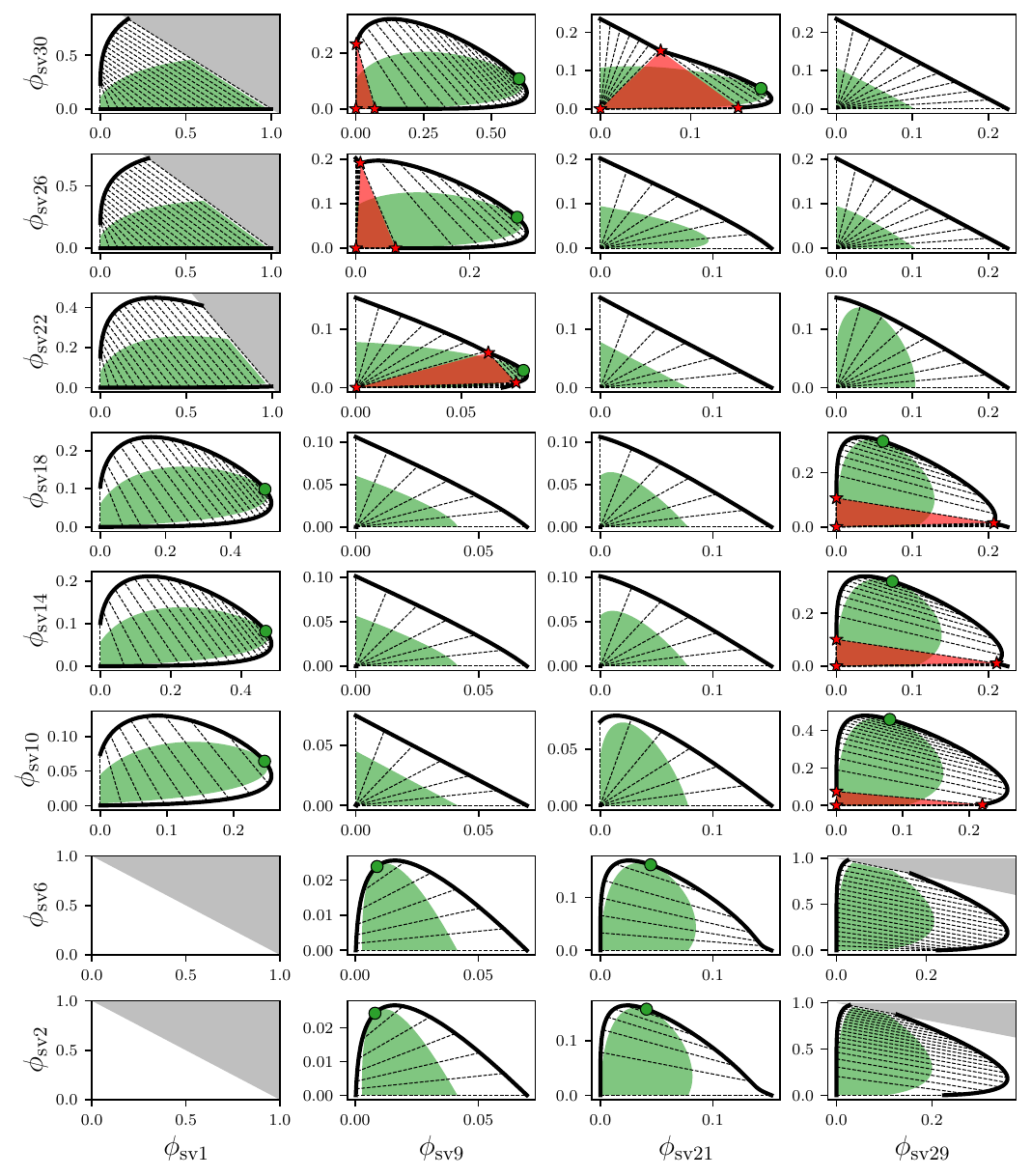}
\vskip -0.4cm
   \caption{RPA-predicted phase diagrams of select sv sequence pairs
from Table~S1. Volume fractions ($\phi$s) are labeled
by subscripts of sequence names.
All phase diagrams in this figure are computed at
Bjerrum length $l_{\rm B}=0.8b$ (reduced temperature $T^*=b/l_{\rm B}=1.25$), 
and no electrostatic screening (inverse Debye screening length 
$\kappa_{\rm D}=0$), i.e., same conditions as those for some of the RPA and 
phase-field results in the maintext (Figs.~1f-j, 3, and 5--7).
Forbidden areas with total volume fraction $>1$ are 
in grey; otherwise the notation is identical to that in maintext 
Fig.1f-j, viz., spinodal and tenary LLPS regimes are in 
green and red, respectively, ternary coexisting phases are marked by 
red stars, binodal phase boundaries are represented by solid black curves, 
binary tielines are dashed, and critical points are marked by 
green circles.
}
   \label{figS1}
\end{figure}
\vfill\eject

\begin{figure}[t]
   \centering
   \includegraphics[width=0.90\columnwidth]{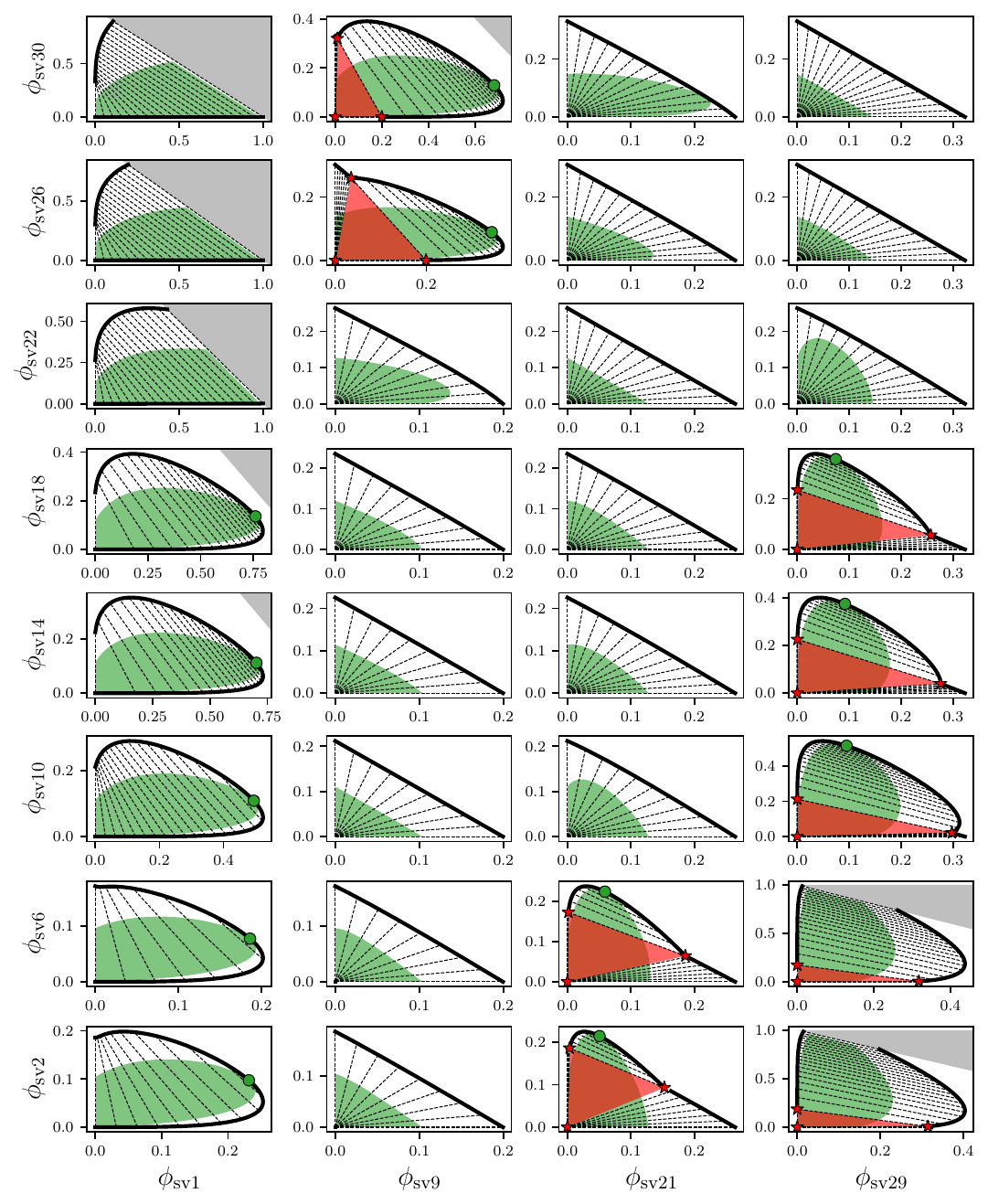}
   \caption{RPA-predicted phase diagrams of sv sequences at a
lower temperature. Same as Fig.~S1 but at a larger 
$l_{\rm B}=2.0b$ corresponding to a lower $T^*=0.5$.
}
   \label{figS2}
\end{figure}
\vfill\eject

\begin{figure}[t]
   \centering
   \includegraphics[width=\columnwidth]{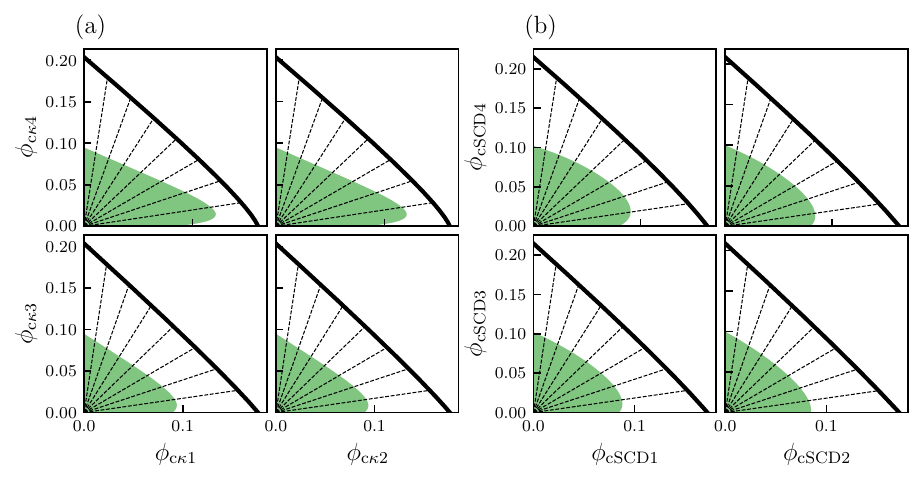}
\vskip -0.4cm
   \caption{RPA-predicted phase diagrams of pairs of
``constant-$\kappa$'' (c$\kappa$) and ``constant-SCD'' (cSCD) sequences 
Pal et al.\cite{Paletal_JPCL2024}
at $l_{\rm B}=0.8b$ ($T^*=b/l_{\rm B}=0.125$), 
and $\kappa_{\rm D}=0$. The notation is the
same as that for Figs.~S1 and S2.
(a) Sequence pairs with 
essentially constant $\kappa\approx 0.5$ but with either
low $-{\rm SCD}$ (c$\kappa$1, c$\kappa$2)
or high $-{\rm SCD}$ (c$\kappa$3, c$\kappa$4).
(b) Sequence pairs with 
essentially constant
$-{\rm SCD}\approx 10$ but with either
low $\kappa$ (cSCD1, cSCD2) or high $\kappa$ (cSCD3, cSCD4).
}
   \label{figS3}
\end{figure}
\vfill\eject

\begin{figure}[t]
   \centering
   \includegraphics[width=\columnwidth]{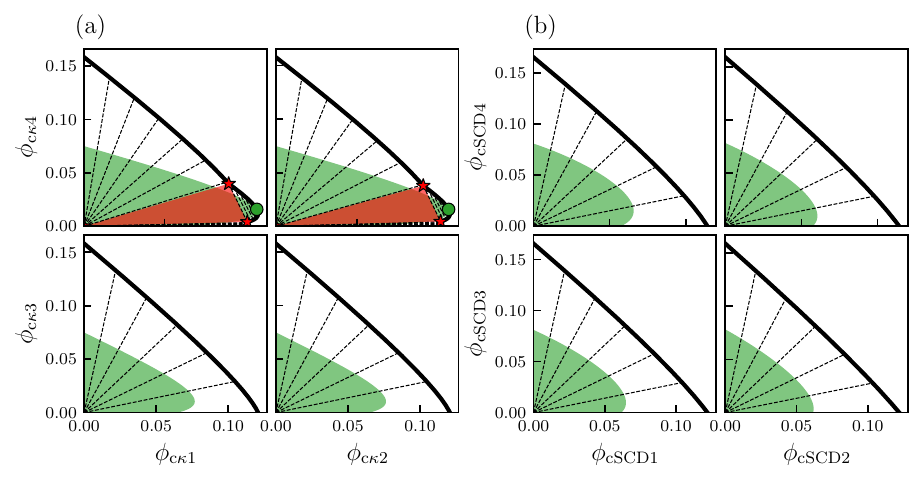}
\vskip -0.4cm
   \caption{Some of the c$\kappa$ sequence pairs exhibit ternary LLPS
at a higher temperature. Same as Fig.~S3 but
for $l_{\rm B}=0.5b$ ($T^*=b/l_{\rm B}=0.2$), $\kappa_{\rm D}=0$. Ternary
LLPS is observed for the two sequence pairs
both involving c$\kappa$4 [top row of (a)].
}
   \label{figS4}
\end{figure}
\vfill\eject

\begin{figure}[t]
   \centering
   \includegraphics[width=\columnwidth]{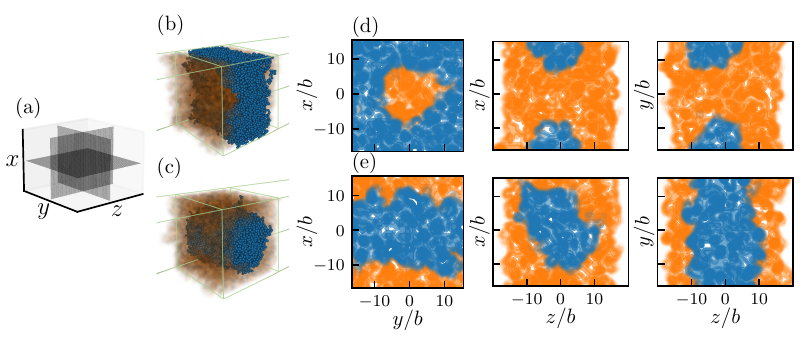}
   \caption{Periodic boundary conditions in MD simulation can lead to
complex tube-like organizations of condensates with demixed phases.
Examples here are for the sv1 (golden)--sv28 (blue) sequence pair.
(a) The planes for the cross-sectional views in (d).
(b,c) 3D view of a condensed droplet with 
chain numbers (b) $(n_{\rm sv1},n_{\rm sv28})=(325,175)$ or (c) $(350,150)$;
(c) is a 3D view of the top snapshot in maintext 
Fig.~2d. As in maintext Fig.~2c,
sv28 beads are depicted explicitly whereas sv1 bead density is
rendered translucent.
(d) Top: The cross-sectional view for the 
$(n_{\rm sv1},n_{\rm sv28})=(325,175)$ snapshot
shows a sv1-rich tube in the $z$-direction. 
Bottom: The cross-sectional view for 
the $(n_{\rm sv1},n_{\rm sv28})=(350,150)$ snapshot
shows an sv28-rich tube in the $y$ direction. 
}
   \label{figS5}
\end{figure}
\vfill\eject


\begin{figure}[t]
   \centering
   \includegraphics[width=\columnwidth]{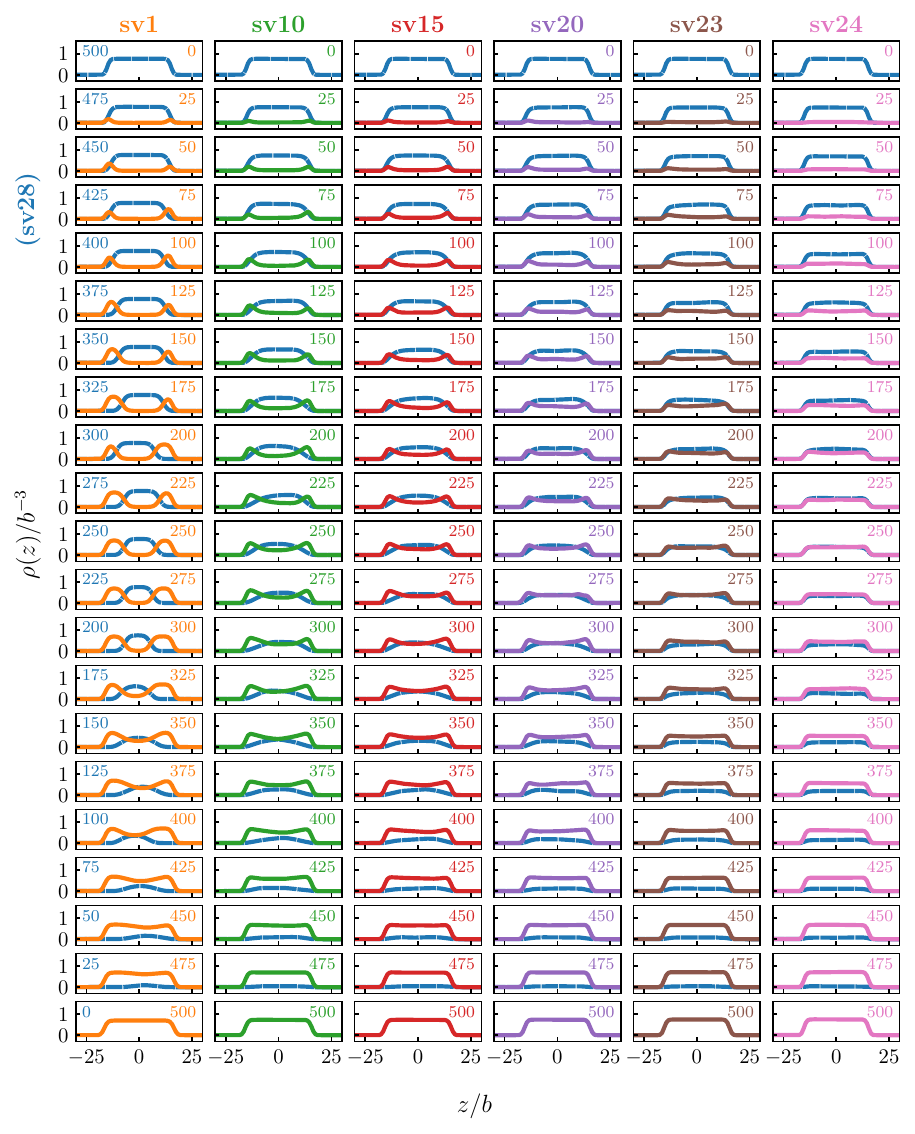}
\vskip -0.4cm
   \caption{MD-simulated 1D density profiles
at $T^*=0.65$ with different sequence proportions
indicated by the numbers of chains simulated
(same color code as that in maintext Fig.~2); $z$ is the
elongated dimension of the simulation box. Bead density $\rho(z)$
is the numbers of a given sequence's chain beads in
successive slabs of thickness $b$ along $z$ divided by the
volume of the slab in units of $b^3$.
}
   \label{fig3}
\end{figure}
\vfill\eject


\begin{figure}[t]
   \centering
   \includegraphics[width=\columnwidth]{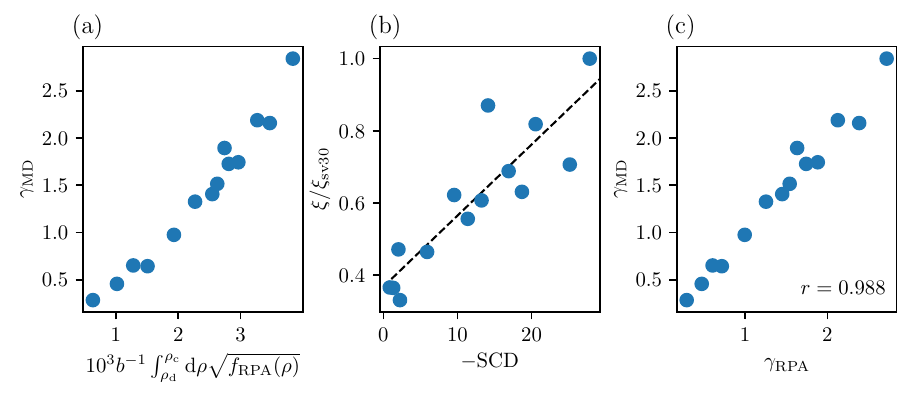}
\vskip -0.4cm
   \caption{Capturing surface tension effects 
in coarse-grained MD simulations of polyampholyte condensates by
RPA-based phase-field theory.
(a) The relationship between the surface tension $\gamma_{\rm MD}$ values 
obtained previously by Devajaran et al.\cite{MittalNatComm2024} for 
50-bead polyampholytic sv sequences on one hand and the phase-field
quantity $\int_{\rho_{\rm d}}^{\rho_{\rm c}} {\rm d}\rho 
\sqrt{f_{\rm RPA}(\rho)}$ on the other 
provides the $\xi$ parameters that penalize density
gradient for the sv sequences in our phase-field theory. 
(b) The resulting $\xi$
values (in units of $\xi_{\rm sv30}$ for sequence sv30) 
vary approximately linearly with the sequences' $-$SCD parameters.
(c) Applying the linear relationship in (b) yields 
$\gamma_{\rm RPA}$s in the phase-field theory, which
correlate very well with the $\gamma_{\rm MD}$s inferred
from explicit-chain MD simulations. 
}
   \label{figS6}
\end{figure}
\vfill\eject

\begin{figure}[t]
   \centering
   \includegraphics[width=\columnwidth]{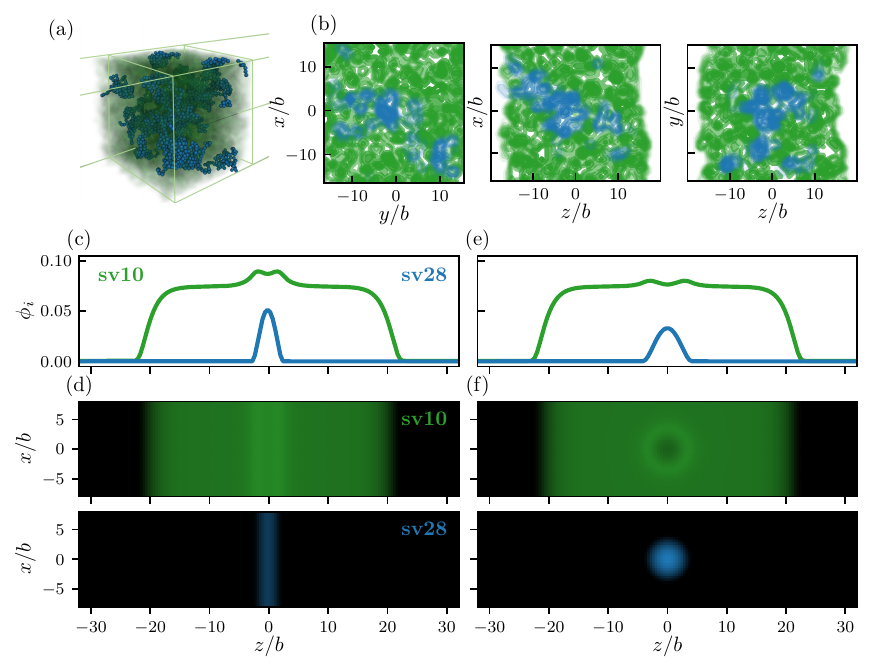}
\vskip -0.4cm
   \caption{Effects of surface tension on condensates
in MD and phase-field theory. Results are shown for the sv10--sv28 pair.
(a,b) A MD snapshot of a condensed droplet with 450 sv10 chains (green)
and 50 sv28 chains (blue) in a simulation box with periodic boundary 
conditions, depicted in the same perspective (a) and cross-sectional views (b) 
as those in Fig.~S5b--d for another sequence pair.
The corresponding $\rho(z)$ averaged over multiple MD snapshots 
is given in Fig.~S6 (2nd column, 3rd panel from bottom).
(c) One-dimensional $z$-dependent volume fraction ($\phi_i$) profiles computed
using 1D phase-field theory.
(d) The same $\phi_i$ profiles, which depends only on $z$ but not on $x$, are 
depicted by heat maps for individual sequence species (same color code). Areas
with very low polyampholyte volume fraction are in black.
(e) 1D $z$-dependent $\phi_i$ profiles computed 
using 2D phase-field theory, wherein
$x$-dependence is averaged over.
(f) The heat maps provides the full 2D $(x,z)$-dependent $\phi_i$ profiles 
from 2D phase-field theory. Numerical results of both the 1D and 2D phase-field 
theories are obtained from the formulation described in the maintext 
and SI text, with $L_z = 64b$, $l_{\rm B}=0.8b$ 
($T^*=b/l_{\rm B}=1.25$), $\kappa_{\rm D} = 0$, $\rho_{\rm tot} = 5b^{-3}$, 
$\phi_{\rm sv10}= 0.0477$, and $\phi_{\rm sv28} = 0.00231$. 
The 1D and 2D phase-field results follow from assuming, respectively, 
$\phi(x,y,x)=\phi(z)$ (no $x$ or $y$ dependence) and $\phi(x,y,z)=\phi(x,z)$.  
In our numerical calculation, the numbers of lattice points used for
the 1D case is $N_z = 150$, those for the 2D case are
$N_z = 148$ and $N_x = 37$ with $L_x = 16b$.  
}
   \label{figS7}
\end{figure}
\vfill\eject

\begin{figure}[t]
   \centering
   \includegraphics[width=\columnwidth]{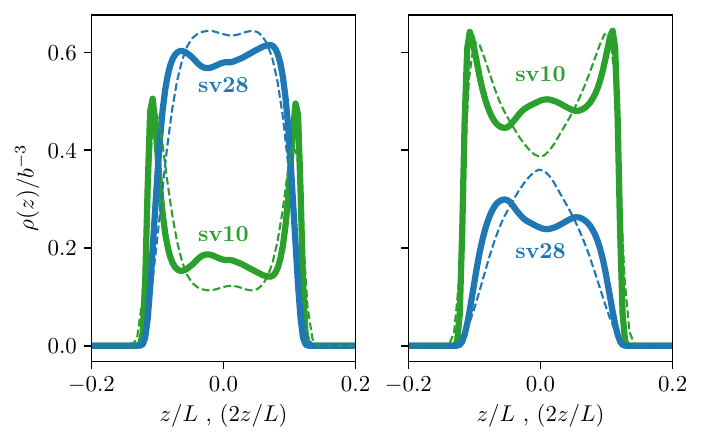}
\vskip -0.4cm
   \caption{MD-simulated density profiles $\rho(z)/b^{-3}$
(similar to those
in Fig.~S6) for two sv10-sv28 systems each containing 
1,000 chains total (solid curves) Left: 300 sv10 and 700 sv28 chains; 
right: 700 sv10 and 300 sv28 chains. These profiles are contrast with 
those simulated under the same conditions and the same 
sv10/sv28 proportions but each containing 500 chains total (dashed curves). 
The latter profiles
are equivalent to the 150-350 sv10-sv28 (left) and 350-150 sv10-sv28 (right) 
profiles in Fig.~S6. The dimensions of the final simulation box for all
four systems is $33\times 33\times 264b^3$. Here the unit for the
horizontal scale is $L=264b$. To facilitate comparison of surface tension
effects on the overall droplet configuration, the 500-chain profiles 
(dashed) are plotted in an expanded horizontal scale ($2z/L$) double that 
for the 1,000-chain profiles ($z/L$) such that the horizontal extents 
of all condensed droplets appear about equal in the plots.
}
   \label{figS8}
\end{figure}
\vfill\eject

\begin{figure}[t]
   \centering
   \includegraphics[width=0.92\columnwidth]{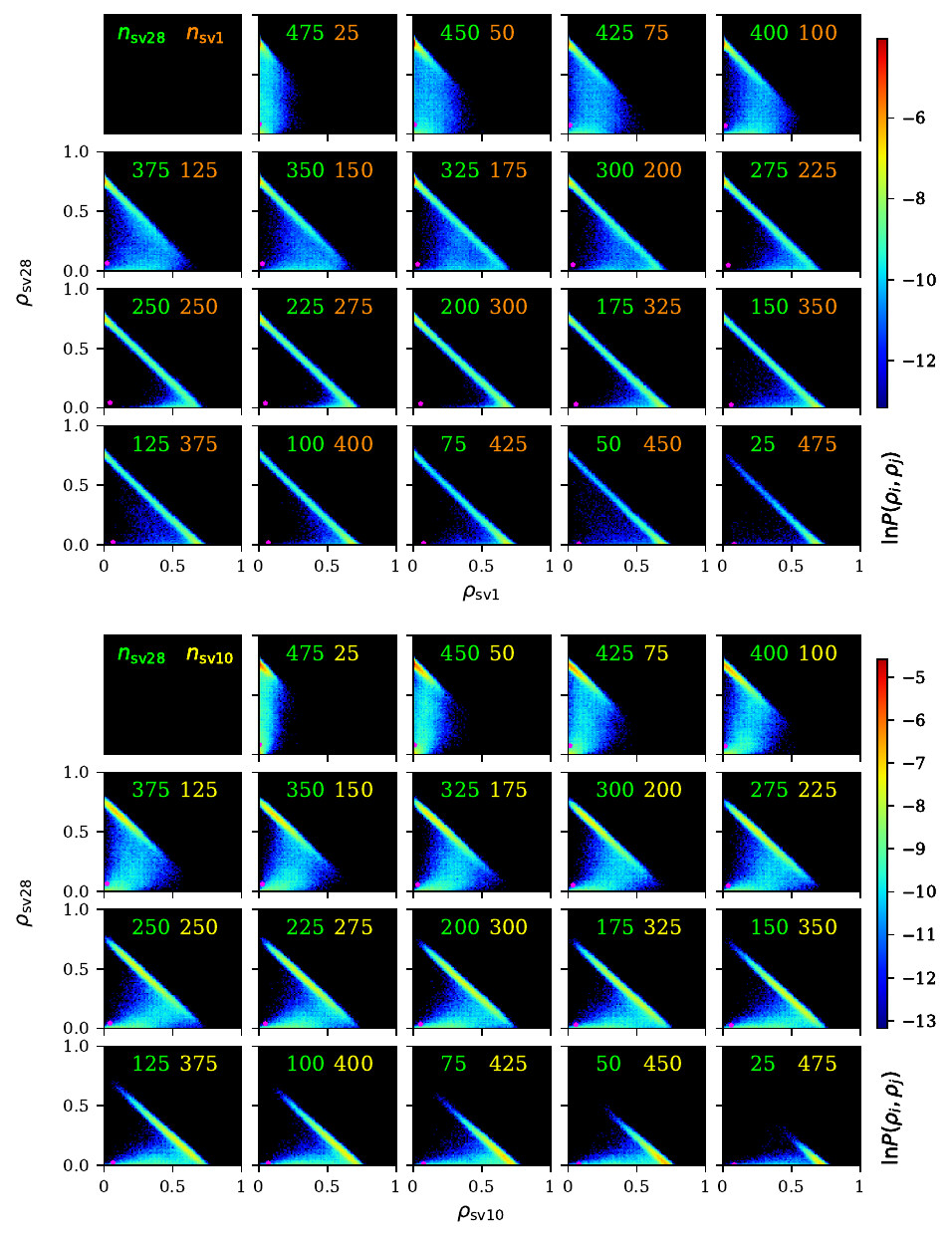}
\vskip -0.4cm
   \caption{MD-simulated density distributions.
Normalized distributions $P(\rho_i,\rho_j)$ of local bead densities
$\rho_i,\rho_j$ for $i$--$j$ $=$ sv1--sv28 (top) and sv10--sv28 (bottom) 
are obtained as in maintext 
Fig.~4. The heat maps show logarithmic distributions (right scales, 
$P=0$ depicted as black) for different numbers of chains $(n_i,n_j)$ 
for the two 
sequences in the MD simulation as indicated by the $n$ values in 
each map in accordance with the color code in the top-left panel
for the sequence pair. 
}
   \label{figS9}
\end{figure}
\vfill\eject

\begin{figure}[t]
   \centering
   \includegraphics[width=0.92\columnwidth]{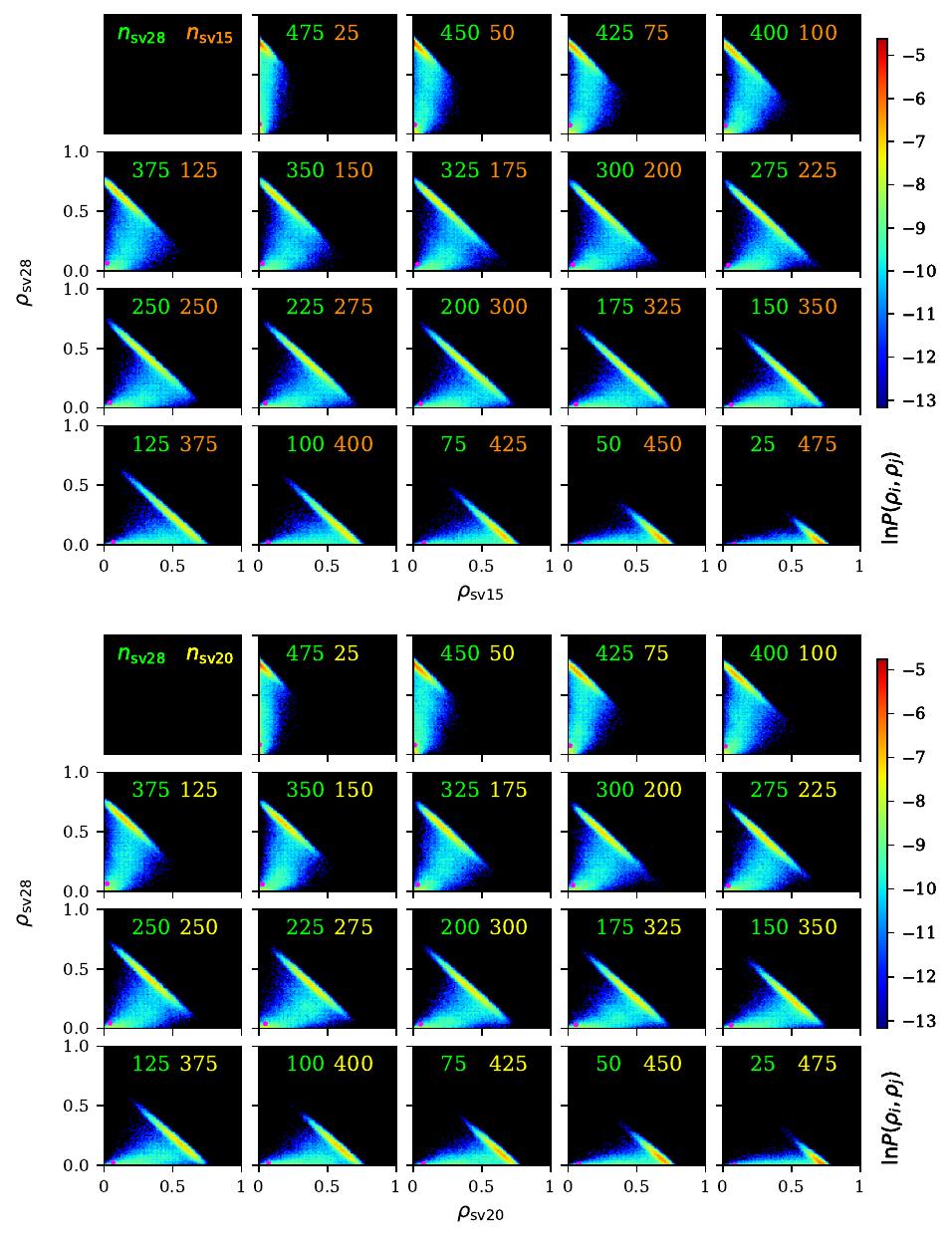}
\vskip -0.4cm
   \caption{MD-simulated distributions of sequence pair densities.
Same as Fig.~S10 but for the sequence pairs sv15--sv28 (top) and
sv20--sv28 (bottom).
}
   \label{figS10}
\end{figure}
\vfill\eject

\begin{figure}[t]
   \centering
   \includegraphics[width=0.92\columnwidth]{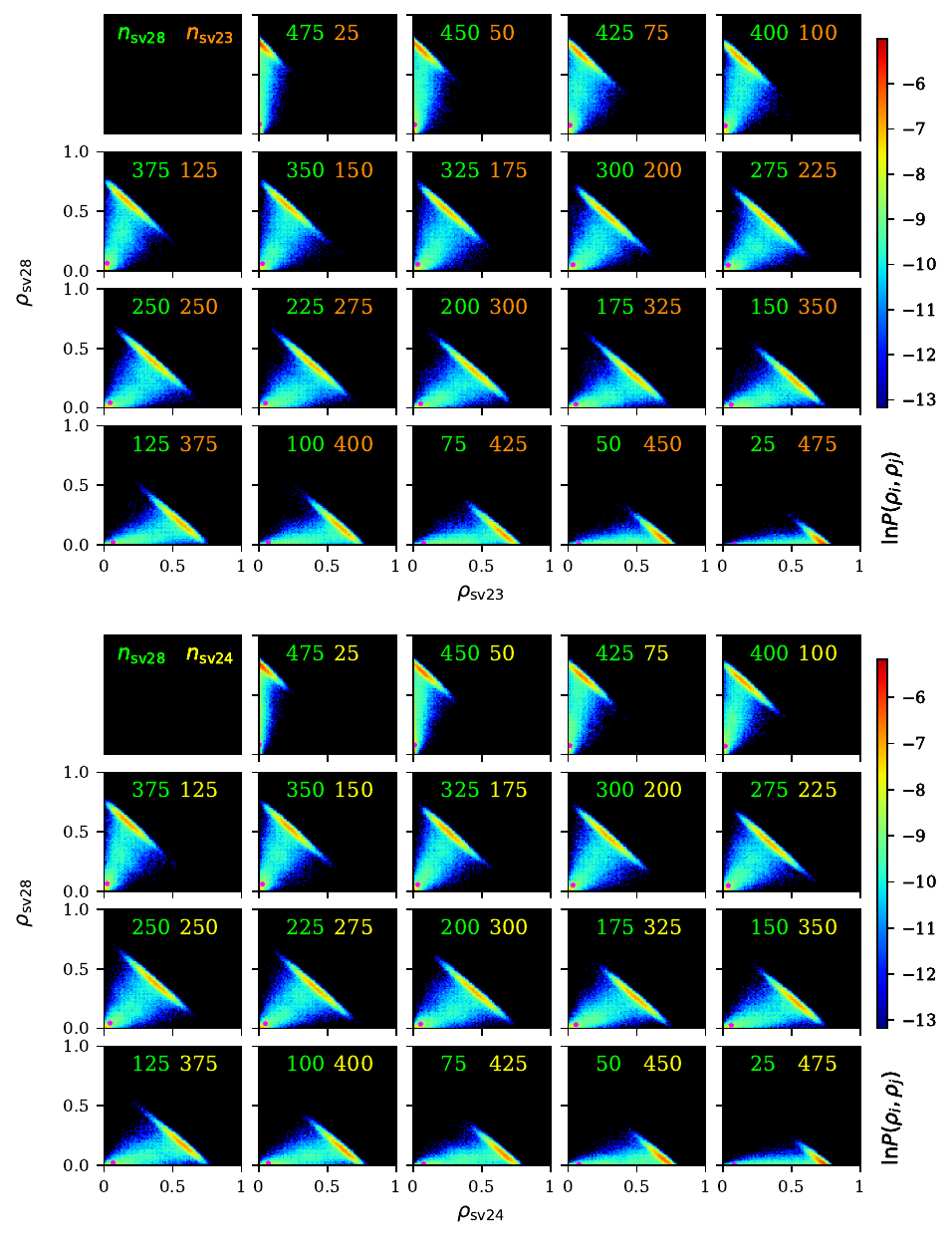}
\vskip -0.4cm
   \caption{MD-simulated distributions of sequence pair densities
Same as Fig.~S10 but for the sequence pairs sv23--sv28 (top) and
sv24--sv28 (bottom).
}
   \label{figS11}
\end{figure}
\vfill\eject

\begin{figure}[t]
   \centering
   \includegraphics[width=\columnwidth]{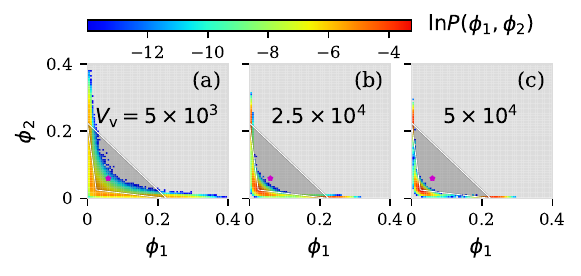}
\vskip -0.4cm
   \caption{System size-dependent phase behavior of a Flory-Huggins (FH) system
that exhibits ternary LLPS at infinite system size. 
Shown results are from a Gibbs ensemble-like Monte Carlo (MC) model 
a system with total 
volume $V$ formulated similarly to those applied for Fig.~7 in the maintext. 
Here 6,000 MC samples is used; $V=200V_{\rm v}$ where $V_{\rm v}$ is voxel 
volume; and $\phi_1$ and $\phi_2$ are volume fractions of two solute species.
The underlying FH free energy is taken from that used for Fig.~7 
of Lin et al.\cite{njp2017}, viz., chain lengths $N_1=N_2=50$, FH interaction 
parameters $\chi_{11}=\chi_{22}=0.66$, and $\chi_{12}=0.33$.
The heat maps show $P(\phi_1,\phi_2)>0$
distributions in accordance with the logarithmic scale at the top, 
whereas $P(\phi_1,\phi_2)=0$ is indicated by light gray.
The dark gray triangles are the FH-predicted ternary LLPS
areas for infinite $V$. The purple star in each panel marks
the input volume fractions $(\bar{\phi}_1,\bar{\phi}_2)=(0.059,0.059)$. 
}
   \label{figS12}
\end{figure}
\vfill\eject

\begin{figure}[t]
   \centering
\vskip -0.4cm
   \includegraphics[width=\columnwidth]{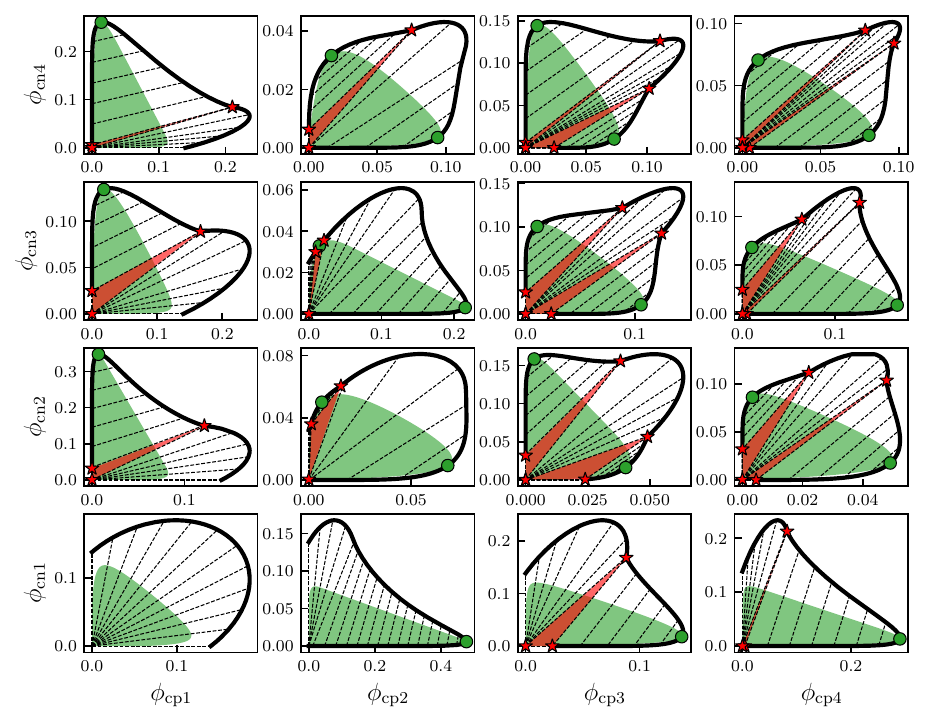}
   \caption{RPA-predicted phase diagrams of pairs of sequences with
net positive or negative charges. The charged sequences 
are given in Table~S1. The phase diagrams are computed at 
$l_{\rm B}=2b$ ($T^*=0.5$) with Debye screening ($\kappa_{\rm D}=0.5b^{-1}$).
The notation for the phase diagrams
are the same as that for Figs.~S1--4. Ternary LLPS regimes
(red areas) are seen in all the phase diagrams except for sequence
pairs cn1--cp1 and cn1--cp2.
}
   \label{figS13}
\end{figure}
\vfill\eject




\vfill\eject
\clearpage
\clearpage

\noindent
{\Large\bf References}\\

\clearpage
\end{document}